\documentclass[a4paper,11pt]{article}
\pdfoutput=1 

\usepackage{PRD} 

\usepackage{commath}
\usepackage{amsthm,amssymb,amsmath,mathrsfs}
\usepackage{feynmf}
\usepackage[T1]{fontenc} 
\usepackage{mathtools}
\usepackage{fixmath}

\DeclarePairedDelimiterX\braket[2]{\langle}{\rangle}{#1 \delimsize\vert #2}
\usepackage{graphicx}
\usepackage{graphicx}
\graphicspath{ {images/} }

\def\be {\begin{equation}}
\def\ee {\end{equation}}
\def\bea {\begin{eqnarray}}
\def\eea {\end{eqnarray}}
\def\bc {\begin{center}}
\def\ec {\end{center}}
\def\bfg {\begin{figure}}
\def\efg {\end{figure}}
\def\bi {\begin{itemize}}
\def\ei {\end{itemize}}

\def\le {\left}
\def\ri {\right}
\def\p {\partial}

\def\a  {\alpha}
\def\b  {\beta}
\def\g  {\gamma}

\def\e  {\eta}
\def\m  {\mu}
\def\n  {\nu}

\def\w {\omega}
\def\dw{\p_\omega}
\def\du{\p_u}
\def\s {\sigma}

\title{Modified Wigner equations and continuous spin gauge field
}

\author
{Mojtaba Najafizadeh}
\affiliation[a]
{ Laboratoire de Math\'ematiques et Physique Th\'eorique\\
Unit\'e Mixte de Recherche $7350$ du CNRS\\
F\'ed\'eration de Recherche $2964$ Denis Poisson\\
Universit\'e Fran\c{c}ois Rabelais, Parc de Grandmont\\
37200 Tours, France}

\affiliation[b]
{School of Physics\\
Institute for Research in Fundamental Sciences (IPM)\\
P.O.Box 19395-5531, Tehran, Iran}

\emailAdd{mojtaba.najafizadeh@lmpt.univ-tours.fr}

\abstract{In this paper, we first propose the bosonic (fermionic) modified Wigner equations for continuous spin particle (CSP)\,. Secondly, starting from the (Fang-)Fronsdal-like equation, we will reach to the modified action of bosonic (fermionic) continuous spin gauge field in flat spacetime, presented recently by Metsaev in A(dS) spacetime\,. We shall also explain how to obtain the proposed modified Wigner equations from the gauge-fixed equations of motion\,. Finally, we will consider the massive bosonic (fermionic) higher-spin action and, by taking the infinite spin limit, we will arrive at the modified bosonic (fermionic) CSP action\,.}


\begin{document}

\maketitle



\section{Introduction}
Elementary particles propagating on Minkowski spacetime were classified by Wigner using the unitary irreducible representations of the Poincar\'e group $ISO(3,1)$\,\cite{Wigner}\,(see also \cite{Unitary} for more details in any dimension). In D spacetime dimensions, the massive particles are determined by representations of the rotation group $SO(D - 1)$\,. For massless particles there are two different representations; the familiar massless particles (helicity particles) which describe particles with a finite number of degrees of freedom determined by representations of the Euclidean group $E_{D-2}=ISO(D - 2)$, and the less-familiar massless particle (continuous spin particle (CSP)) which describes a particle with an infinite number of physical degrees of freedom per spacetime point characterized by the representations of the short little group $SO(D - 3)$, the little group of $E_{D - 2}$\,\cite{Brink:2002zx}\,. For the both massless representations, the eigenvalue of the quadratic Casimir operator $C_2 := P^2 $ (the square of the momentum $P_\m$) vanishes\,. For the helicity representation, the eigenvalue of the quartic Casimir operator $C_4 := W^2$ (the square of the Pauli-Lubanski vector ${W}^\m= \frac{1}{2} \, \epsilon^{\m\n \rho\sigma} \, {P}_\n \, {J}_{\rho\sigma}$) is zero, while the one for the continuous spin representation becomes $\m^2$ (a positive real parameter)\,.

From the group theoretical point of view, Khan and Ramond have illustrated that the continuous spin representation can be obtained from the massive higher-spin representation by taking a limit; involving the mass going to zero and the spin to infinity with their product being fixed \cite{KR}\,. Using this fact, the equation of motion for the single and double valued continuous spin representations were firstly found in \cite{BM} by taking the limit from the massive higher-spin field equations and are known as the so-called ``Fronsdal-like'' and ``Fang-Fronsdal-like'' equations with the trace-like constraints on the gauge fields and parameters. Although these equations recover the Fronsdal \cite{F 1} and Fang-Fronsdal \cite{F 2} equations with the trace constraints on the gauge fields and parameters, they can not be obtained from an action principle.

Later on, the first action principle for a single bosonic continuous spin gauge field, including unconstrained gauge field and parameter, was proposed by Schuster and Toro in \cite{ST PRD}\,(see also their earlier works \cite{Schuster:2013pxj}-\cite{Schuster:2013pta} and related remarks in \cite{Rivelles:2014fsa, Rivelles:2016})\,. The authors also found a constrained formulation of the CSP action in tensor form which was rank-mixing for non-zero $\m$\, (but rank-diagonal for $\m=0$). Afterwards, a similar action principle for the unconstrained fermionic CSP field was found in \cite{BNS}\,. By performing a Fourier transformation in the auxiliary vector, it was shown the acquired equations of motion in \cite{ST PRD, BNS} are equivalent to the Fronsdal-like and Fang-Fronsdal-like equations\,. It is shown, by solving the trace-like constraint in the Fourier space, the bosonic (fermionic) CSP action can be obtained from the (Fang-)Fronsdal-like equation \cite{BMN}\,.

Recently, the modified action principles for the bosonic \cite{Metsaev: B CSP} and fermionic \cite{Metsaev: F CSP} continuous spin gauge fields, comprising the constrained gauge field and parameter, were found by Metsaev in the Minkowski and (anti) de Sitter spacetimes for any dimension\,. Modified de Donder gauge conditions which simplify the equations of motion were found and it was demonstrated that partition functions of the bosonic and fermionic continuous spin gauge fields are equal to one\,. All above formulations for the continuous spin gauge field were constructed in the metric-like approach, while the frame-like formulation in 3-dimensions was recently found in \cite{Zinoviev:2017rnj}\,.

To our knowledge, so far the modified Wigner equations have not been discussed previously in the literature. In addition, it is interesting to examine the relationship between the modified CSP action \cite{Metsaev: B CSP, Metsaev: F CSP} and the (Fang-)Fronsdal-like equation \cite{BM} using a field redefinition in D dimensions, as well as the connection between the constrained formalism of the Schuster-Toro action \cite{ST PRD} and the Metsaev action \cite{Metsaev: B CSP} in flat spacetime\,. Moreover, obtaining the modified CSP actions \cite{Metsaev: B CSP, Metsaev: F CSP} by taking a limit of the massive higher-spin actions \cite{Metsaev:2008fs, Metsaev:2006zy} are attractive\,. These are the key issues to discuss in this paper\,.

The layout of the letter is as follows. In section \ref{MWE}\,, using a field redefinition, we obtain the modified Wigner equations for bosons and fermions. In section \ref{MBCSP}\,, we consider the Fronsdal-like equation and, by applying a field redefinition, acquire the modified bosonic CSP action in the Minkowski spacetime, presented by Metsaev in (A)dS spacetime \cite{Metsaev: B CSP}\,. We also find that our modified bosonic Wigner equations can be obtained from the gauge-fixed equation of motion. In addition, we make a relation between the modified bosonic CSP action and the tensor action of Schuster and Toro in \cite{ST PRD}\,. In section \ref{MFCSP}\,, we follow a similar procedure, as the section \ref{MBCSP}\,, for fermions\,. In the section \ref{LLbar}\,, we consider the massive bosonic \cite{Metsaev:2008fs} and fermionic \cite{Metsaev:2006zy} higher-spin actions in flat spacetime and, by taking the infinite spin limit, arrive at the modified bosonic and fermionic CSP actions\,. We conclude and discuss in section \ref{s8}\,. In appendices; we present our conventions in the appendix \ref{conv.}\,. Transformation operators which convert the trace-like constraints to the trace ones are introduced in the appendix \ref{Trans. oper.}\,. How to obtain the modified Wigner equations will be explained in the appendix \ref{MWEss}\,. We have computed useful relations in the appendix \ref{useful} which will be used in the sections \ref{MBCSP} and \ref{MFCSP}\,.


\section{Modified Wigner equations} \label{MWE}
 For the both kind of CSPs (bosons and fermions), wave equations describing a single CSP were found by Wigner in \cite{Wigner}\,. It was shown these equations can be obtained from the Fronsdal-like and Fang-Fronsdal-like equations after fixing the gauge field \cite{BM}\,. In this section, using a field redefinition, we will obtain the bosonic and fermionic modified Wigner equations which is expected to arise from the modified bosonic \cite{Metsaev: B CSP} and fermionic \cite{Metsaev: F CSP} CSP actions in flat spacetime\,. We shall present these equations separately for bosons and fermions in the following subsections\,.

\subsection{Bosonic equations} \label{B C Four}

The bosonic Wigner equations in D-dimensional spacetime are given by
 \bea
\le(\,p\cdot \p_\e \,-\,i\,\m \ri)\,\overline{ \boldsymbol{\varphi}}(p,\e)=0\,,&&\label{cspin1e}\\
({\e^2}\,-\,1\,)\,\overline{\boldsymbol{\varphi}}(p,\e)=0\,,\label{cspin2e}&&\\
\left(\,p\cdot{\e}\right)\,\overline{\boldsymbol{\varphi}}(p,\e)=0\,,\label{cspin3e}&&\\
p^2\,\overline{\boldsymbol{\varphi}}(p,\e)=0\,,&& \label{cspin4e}
\eea
where $\e^\m$ is a D-dimensional auxiliary vector and $\m$ is a real parameter with the
dimension of a mass\,. If we define the Pauli-Lubanski vector\,\footnote{\,The above Pauli-Lubanski vector for simplicity is considered in 4-dimensions, while the D-dimensional form can be found in \cite{Brink:2002zx}\,.} ${W}^\m= \frac{1}{2} \, \epsilon^{\m\n \rho\sigma} \, {p}_\n \, {J}_{\rho\sigma}$\,, then its square would be
\be
W^2=2\,(p \cdot \e) (p \cdot \p_\e) (\e \cdot \p_\e) - (p \cdot \e)^2 \, (\p_\e \cdot \p_\e) - \e^2 (p \cdot \p_\e )^2 \,, \label{W 2}
\ee
where $W^2$\, is the Casimir operator of the Poincar\'e algebra and the massless representations can
be labelled by the eigenvalues of this operator
\be
W^2~\overline{\boldsymbol{\varphi}}(p,\e) = \m^2 ~\overline{\boldsymbol{\varphi}}(p,\e)\,. \label{crucial}
\ee
Consequently when $\m=0$ the representations are finite dimensional (helicity representation)\,, and when $\m \neq 0$ the representations are infinite dimensional (continuous spin representation)\,.

In $\w$-space, the Fourier space of $\e$\,, $
\overline{\boldsymbol{\varphi}}(\e)=\int \frac{d^D \e}{(2\,\pi)^{D/2}} ~ e^{-\,i\,{\w \cdot \e}}~\widetilde{\boldsymbol{\varphi}}(\w)
$\,, the Wigner equations \eqref{cspin1e}-\eqref{cspin4e} will take the following forms
\bea
\le(\,p\cdot \w+\,\sigma\,\m \ri)\,\widetilde{ \boldsymbol{\varphi}}(p,\w)=0\,,&&\label{cspin1}\\
\left({\dw \cdot \dw}+\,\sigma\right)\widetilde{\boldsymbol{\varphi}}(p,\w)=0\,,\label{cspin2}&&\\
\left(\,p\cdot{\dw}\right)\widetilde{\boldsymbol{\varphi}}(p,\w)=0\,,\label{cspin3}&&\\
p^2\,\widetilde{\boldsymbol{\varphi}}(p,\w)=0\,,&& \label{cspin4}
\eea
where we have added a $\s$ parameter
\be
\sigma = \left\{
           \begin{array}{ll}
             0, & \hbox{for higher-spin;} \\
             1, & \hbox{for continuous-spin.}
           \end{array}
         \right.  \label{sigma}
\ee
into the equations to include the similar obtained equations in \cite{BM}\,, describing a massless higher-spin particle \footnote{\,In the rest of paper, we shall keep $\s$ parameter in our formulations in order to read the higher-spin results simultaneously\,.}\,. The equation \eqref{cspin2} express that the wave function is traceless when $\s=0$\,, while it is traceless-like (trace-full) when $\s=1$\,. Here, our goal is to have a traceless field for both values of $\s$\,. Therefore, for this purpose, we define
\be
\widetilde{\boldsymbol{\varphi}}(p,\w) = \mathbf{P}_\varepsilon ~\boldsymbol{\varphi}(p,\w)\,,
\label{P_eps 1}
\ee
where $\boldsymbol{\varphi}(p,\w)$ is a redefined gauge field and the operator
\be
\mathbf{P}_\varepsilon = ~ \sum_{n=0}^{\infty}~\sigma^n~ \omega^{\,2n}~ \frac{(-1)^n}{~2^{\,2n} ~ n!~ (N+\tfrac{D}{2})_n~} \,. \label{P_eps 2}
\ee
In this definition, D is the dimension of spacetime, $N={\w\cdot\dw}$\, and $(a)_n$ is the rising Pochhammer symbol \eqref{Pochhammer}\,. Note that in the higher-spin limit ($\s=0$)\,, the operator $\mathbf{P}_\varepsilon=1$ and the redefined $\boldsymbol{\varphi}$ and non-redefined $\widetilde{\boldsymbol{\varphi}}$ fields are equal to each other\,. Applying \eqref{P_eps 1} into \eqref{cspin2}, and referring to the appendix \ref{MWEss}, we find that the equation \eqref{cspin2} can be reduces to $\left({\dw \cdot \dw}\right){\boldsymbol{\varphi}}=0$\,, which is a tracelessness condition for a redefined (modified) CSP field\,. Plugging \eqref{P_eps 1} into the other Wigner equations \eqref{cspin1}, \eqref{cspin3} and \eqref{cspin4}, we finally find (see the appendix \ref{MWEss})
\bea
\le(\,p\cdot \w\,+\,\s\,\m \,+\,\s\,\m\,\w^2\, \frac{1}{(2N+D)(2N+D-2)} \, \ri){\boldsymbol{\varphi}}(p,\w)=0\,,~~&&\label{cspin11}\\
\Big({\dw \cdot \dw}\Big)~{\boldsymbol{\varphi}}(p,\w)=0\,,~~\label{cspin22}&&\\
\left(\,p\cdot\dw\,+\,\s\,\m\,\frac{1}{2N+D-2}\,\right){\boldsymbol{\varphi}}(p,\w)=0\,,~~\label{cspin33}&&\\
p^2\,{\boldsymbol{\varphi}}(p,\w)=0\,,~~&& \label{cspin44}
\eea
which we will call them ``modified Wigner equations'' for bosons\,. A surprising equation in above is \eqref{cspin33} which demonstrates the modified CSP field is not divergence-less\,! Let us to discuss a bit about this issue in the conclusions\,. We shall see later (in the section \ref{MBCSP})
these modified equations can be extracted from the modified CSP action \cite{Metsaev: B CSP} by fixing the gauge field.

In the Wigner equations, the last two equations can be obtained from the first two independent ones \cite{BM}\,. This fact can be also examined for the modified Wigner equations by writing \eqref{cspin11} and \eqref{cspin22}, respectively, as
\be
 {\cal O}_1\,\boldsymbol{\varphi} = 0\,, \quad \quad\quad\quad  {\cal O}_2\,\boldsymbol{\varphi} = 0 \,.
\ee
Then the equations \eqref{cspin33} and \eqref{cspin44} can be read from
\bea
  \Big[ \,  {\cal O}_1 \,,\, {\cal O}_2 \, \Big]~\boldsymbol{\varphi} \, &=&\, -\,2\, {\cal O}_3  \,\boldsymbol{\varphi} \,
= \,-\,2\, \Big(\,p\cdot\dw\,+
\,\s\,\m\,\frac{1}{{2N+D-2}}\,\Big) \,\boldsymbol{\varphi} \,=\,0 \,,    \\
 \Big[ \,  {\cal O}_1 \,, \, {\cal O}_3  \, \Big]~\boldsymbol{\varphi} \,
 &=& \, -\, {\cal O}_4\, {\boldsymbol{\varphi}} \,=\,-\, p^2 \, {\boldsymbol{\varphi}}=0\,,
\eea
as
\be
 {\cal O}_3\,\boldsymbol{\varphi} = 0\,, \quad \quad\quad\quad  {\cal O}_4\,\boldsymbol{\varphi} = 0 \,.
\ee

In $\e$-space, the Fourier transform of $\w$\,, the modified Wigner equations \eqref{cspin11}-\eqref{cspin44} become
\bea
\le(\,p\cdot \p_\e\,-\,i\,\s\,\m \,+\,i\,\s\,\m\,(\p_\e \cdot \p_\e)\, \frac{1}{(2N+D)(2N+D+2)} \, \ri){\widehat{\boldsymbol{\varphi}}}(p,\e)=0\,,~~&&\label{cspin11e}\\
\e^2 \,{\widehat{\boldsymbol{\varphi}}}(p,\e)=0\,,~~\label{cspin22e}&&\\
\left(\,p\cdot\e\,+\,i\,\s\,\m\,\frac{1}{2N+D+2}\,\right){\widehat{\boldsymbol{\varphi}}}(p,\e)=0\,,~~\label{cspin33e}&&\\
p^2\,{\widehat{\boldsymbol{\varphi}}}(p,\e)=0\,,~~&& \label{cspin44e}
\eea
where here $N={\e \cdot \p_\e}$\,\footnote{\,To lighten notation, we have omitted the subscripts in $N_\w={\w \cdot \dw}$ and $N_\e={\e \cdot \p_\e}$\,.}.
Using these equations, it is convenient to exhibit that the modified continuous spin field satisfies
\be
W^2 ~ {\widehat{\boldsymbol{\varphi}}}(p,\e) = \s\,\m^2 ~ {\widehat{\boldsymbol{\varphi}}}(p,\e)\,.
\label{w2 csp}
\ee
In 4-dimensions, this relation can be simply checked by applying \eqref{W 2} in \eqref{w2 csp}\,. Note that, as it was expected, the modified CSP field in \eqref{w2 csp} obeys a similar relation as \eqref{crucial} for the CSP field\,.

\subsection{Fermionic equations}

In D-dimensional spacetime, the fermionic Wigner equations are given by
\bea
\le(\,p\cdot \w+\,\sigma\,\m \ri)\,\widetilde{ \boldsymbol{\psi}}(p,\w)=0\,,&&\label{fcspin1}\\
\left({\dw \cdot \dw}+\,\sigma\right)\widetilde{\boldsymbol{\psi}}(p,\w)=0\,,\label{fcspin2}&&\\
\left(\,p\cdot{\dw}\right)\widetilde{\boldsymbol{\psi}}(p,\w)=0\,,\label{fcspin3}&&\\
(\gamma \cdot p)\,\widetilde{\boldsymbol{\psi}}(p,\w)=0\,,&& \label{fcspin4}
\eea
which are similar to the bosonic case except the forth equation. For these equations we define
\be
\widetilde{ \boldsymbol{\psi}}(p,\w) = \mathbf{P}_\xi ~\boldsymbol{\psi}(p,\w) \,,
\label{P Kesi}
\ee
where the operator
\be
\mathbf{P}_\xi =\sum_{k=0}^{\infty}\le[(\sigma\,\gamma \cdot \omega)^{2k} + 2ik\, (\sigma\,\gamma \cdot \omega)^{2k-1}\ri]
\frac{(-1)^k}{~2^{\,2k}~k!~(N+\frac{D}{2})_k~} \,.
\label{P Kesi 1}
\ee
In this relation, D stands for the spacetime dimension, $\g^\m$ for the D-dimensional Dirac Gamma matrices, $(a)_k$ for the rising Pochhammer symbol \eqref{Pochhammer} and $N={\w\cdot\dw}$\,. Note that the first sum in \eqref{P Kesi 1} comprises the even powers of $\g\cdot\w$\,, which is precisely equal to \eqref{P_eps 2}\,, while the second sum includes the odd powers\,. Applying \eqref{P Kesi} in the Wigner equations \eqref{fcspin1}-\eqref{fcspin4} we will respectively arrive at (see the appendix \ref{MWEss})
\bea
\le(\,p\cdot \w\,+\,\s\,\m \,+\,i\,\s\,\m\,(\g\cdot \w)\, \tfrac{2}{(2N+D)(2N+D-2)}\,+\,\s\,\m\,\w^2\, \tfrac{1}{(2N+D)^2} \,\ri){\boldsymbol{\psi}}(p,\w)=0\,,~~&&\label{fcspin11}\\
\left({\g \cdot \dw} \,-\,i\,\s\,(\g\cdot\w)(\g\cdot\dw)\,\tfrac{1}{2N+D+2}\,-\,i\,\s\,\tfrac{2}{2N+D+2} \right)\Big(\g\cdot\dw \Big)\,{\boldsymbol{\psi}}(p,\w)=0\,,~~\label{fcspin22}&&\\
\left(\,p\cdot\dw\,+\,\s\,\m\,\tfrac{1}{2N+D}\,\right){\boldsymbol{\psi}}(p,\w)=0\,,~~\label{fcspin33}&&\\
\le(\gamma \cdot p \,+\,i\,\s\,\m\,\tfrac{2}{2N+D-2}\,+\,\s\,\m\,(\g\cdot\w)\,\tfrac{2}{(2N+D)^2}\,\ri){\boldsymbol{\psi}}(p,\w)=0\,,~~&& \label{fcspin44}
\eea
which we will call them ``modified Wigner equations'' for fermions\,. Similar to the bosonic equations, the third equation \eqref{fcspin33} illustrates that the gauge field is divergence-full\,! Concerning to the fourth equation \eqref{fcspin44}\,, by multiplying $\g\cdot p$ to the left and then using \eqref{fcspin11}\,, it is convenient to show that $p^2\,{\boldsymbol{\psi}}(p,\w)=0$\,, which emphasizes the masslessness of the modified fermionic CSP field, as it was expected\,. Relation of these equations with the modified CSP action \cite{Metsaev: F CSP} will examine in the section \ref{MFCSP}\,.

\section{Modified bosonic continuous spin gauge field}  \label{MBCSP}
The modified bosonic CSP action in (A)dS spacetime was presented by Metsaev in \cite{Metsaev: B CSP}\,. The action was written in a formulation similar to the higher-spin action with the double trace constraint on the gauge field and trace constraint on the gauge parameter\,. In \cite{BMN}\, it is shown that, by starting from the Fronsdal-like equation and working in the Fourier space, we can reach to the Schuster-Toro action, however the relation of the Fronsdal-like equation and the modified CSP action was not manifest so far\,. In this section, we are going to make clear this relation by performing a field redefinition in the Fronsdal-like equation which will consequently lead to the modified CSP action\,. The modified gauge transformation will be presented and the modified Wigner equations will be found from the gauge-fixed equation of motion\,.

\subsection{CSP action}

A continuous spin gauge field is defined as
\be
\widetilde{\Phi}(x,\omega)\,=\,\sum_{s=0}^{\infty}~{\Phi}_s(x,\omega)\,=\,\sum_{s=0}^{\infty}~\,\frac{1}{\,s!\,} ~ \omega^{\m_{1}}   \ldots \omega^{\m_{s}} \, \,
{\Phi}_{\m_{1}\ldots\m_{s}}(x) \,,
\label{G Field omega}
\ee
where ${\Phi}_s$ represent for all totally symmetric massless spin-$s$ fields (higher-spin gauge fields) and $\w^\m$ is a D-dimensional auxiliary vector\,. Then\,, the Fronsdal (higher-spin) and Fronsdal-like (continuous-spin) equations can be written as \cite{BM}
\be
\mkern-13mu\widetilde{\mathbf{F}}_\sigma~\widetilde{\Phi}=\le[\,\Box_x - \left(\omega \cdot \p_x + i\, \sigma \m\right) \left(\p_{\omega} \cdot \p_x  \right)
 +
 {\frac{1}{2}}\,(\omega \cdot  \p_x + i \,\sigma \m)^2
\left( \p_{\omega} \cdot \p_\omega + \sigma \right) \ri]\widetilde{\Phi}(x,\omega)=0\,, \label{eom tilde}
\ee
where the parameter $\s$ was introduced in \eqref{sigma}\,.
Note that when $\sigma=0$\,, ${\Phi}_s$ should be substituted in \eqref{eom tilde} instead of $\widetilde{\Phi}$\,.
The double trace-like constraint on the gauge field $\widetilde{\Phi}$ is given by
\be
\left(\, \p_\omega \cdot \p_\omega\, + \,\sigma \, \right)^2 \, \widetilde{\Phi} (x,\omega) = 0\,.  \label{phi constraint}
\ee
Let us now introduce a redefined (modified) gauge field ${\mathbf{\Phi}}$ through
\be
\widetilde{\Phi}(x,\w) = \mathbf{P}_\Phi ~{\mathbf{\Phi}}(x,\w) \,,  \label{tildePhi - Phi}
\ee
where the operator $\mathbf{P}_\Phi$ (with the similar notations for \eqref{P_eps 2}) is defined as
\be
\mathbf{P}_\Phi = ~ \sum_{n=0}^{\infty}~ \sigma^n~\omega^{\,2n}~ \frac{(-1)^n}{~2^{\,2n} ~ n!~ (N+\tfrac{D}{2}-1)_n~}
\label{P_phi text}\,.
\ee
Then, it can be illustrated (see the appendix \ref{Trans. oper.}) that the redefined CSP field is double traceless $ \left( \p_\omega \cdot \p_\omega  \right)^2 \, {\mathbf{\Phi}} (x,\omega) =0$ (similar to the Fronsdal field)\,.
We are now in a position to rewrite the Fronsdal-like equation in terms of the redefined field\,.
To this end, by plugging \eqref{tildePhi - Phi} into \eqref{eom tilde} and applying the relations \eqref{1}-\eqref{3}\,, we shall get
\be
\mathbf{F}_\sigma~{\mathbf{\Phi}}(x,\w) =0 \,,  \label{eom}
\ee
where ${\mathbf{\Phi}}(x,\w)$ is the redefined gauge field and
\bea
\mathbf{F}_\sigma &=& \Box_x - (\omega \cdot \p_x )(\p_\omega \cdot \p_x) + \frac{1}{2}\,   (\omega \cdot \p_x )^2 (\p_\omega \cdot \p_\omega) - i\,\sigma\mu  (\p_\omega \cdot \p_x)  \nonumber \\
&& ~~~~ - i\,\sigma\mu \, \omega^2 \, \frac{1}{(2N+D)(2N+D-2)} \, (\p_\omega \cdot \p_x) - i\,\sigma\mu \, (\omega \cdot \p_x )\,
\frac{1}{(2N+D-2)} \nonumber\\
&& ~~~~ + i\,\sigma\mu  (\omega \cdot \p_x )(\p_\omega \cdot \p_\omega) + i\,\sigma\mu \,\w^2\,(\w\cdot\p_x)\, \frac{1}{(2N+D)(2N+D+2)}\,(\dw\cdot\dw) \nonumber\\
&& ~~~~ - \frac{1}{2}\, \sigma \m^2 \,  (\dw\cdot\dw) - \sigma \m^2 \,\w^2 \, \frac{1}{(2N+D-2)(2N+D+2)}\,(\dw\cdot\dw) \nonumber\\
&&~~~~+  \sigma \m^2 \,\frac{1}{(2N+D-2)} + \sigma \m^2 \,\w^2 \, \frac{1}{(2N+D)(2N+D-2)^2}
+\, \mathcal{O}(\w^4)\,.
 \eea
In above, due to the double trace constraint ${\mathbf{\Phi}}(x,\dw)\,(\w^2)^{\,2}\,=0$\,, the terms comprising the powers of $\w^4$ (or higher) will be vanished at the level of the action\,, and hence we will not consider such terms in the rest of calculations\,.

In comparison to the spin-two case, the operator $\mathbf{F}_\sigma$ is a ``Ricci-like'' operator and by applying the hermitian conjugates \eqref{hermitian conjugates}\, is not hermitian $\mathbf{F}_\sigma^{\,\dag} \neq \mathbf{F}_\sigma$\,, therefore it can not be led to the action\,. To acquire an hermitian operator, leading to the action, we introduce the ``Einstein-like'' operator
\be
\mathbf{K}_\sigma = \le(1-\,\tfrac{1}{4}\, \w^2\,(\dw\cdot\dw) \ri)\mathbf{F}_\sigma \,,
\ee
which after some calculations, and using the commutation relations in \eqref{commutation 1}\,, becomes
\bea
\mathbf{K}_\sigma &=& \le(1-\,\tfrac{1}{4}\, \w^2\,(\dw\cdot\dw) \ri) \Box_x
-
\le( \w\cdot\p_x-\tfrac{1}{2}\,\w^2(\dw\cdot\p_x)\ri)\le(\dw\cdot\p_x-\tfrac{1}{2}\,\w\cdot\p_x(\dw\cdot\dw)\ri) \nonumber\\
&-&\,i\,\sigma\m\,(\dw\cdot\p_x) - \,i\,\sigma\m \,(\w\cdot\p_x)\,\frac{1}{(2N+D-2)} \nonumber\\
&+&\,i\,\sigma\m\,(\w\cdot\p_x)(\dw\cdot\dw) + \,i\,\sigma\m\, \w^2 \, \frac{1}{(2N+D)}\,(\dw\cdot\p_x)\nonumber\\
&-&\,\tfrac{1}{4}\,i\,\sigma\m\,\w^2\,(\dw\cdot\p_x)(\dw\cdot\dw) -\,\tfrac{1}{4}\,i\,\sigma\m\,\w^2\,(\w\cdot\p_x)\,\frac{1}{(2N+D+2)}\,(\dw\cdot\dw)\nonumber\\
&+&\,\sigma\m^2\, \frac{1}{(2N+D-2)}\,+\,\sigma\m^2\,\w^2\, \frac{1}{4(2N+D+2)} \,(\dw\cdot\dw)\nonumber\\
&-&\,\tfrac{1}{2}\,\sigma\m^2\, (\dw\cdot\dw) -\,\tfrac{1}{2}\,\sigma\m^2\,\w^2\,\frac{1}{(2N+D)(2N+D-2)}\,.
\label{K operator}
\eea
We shall see here the above operator is now clearly hermitian  
when $\sigma=0$\,, while for $\sigma=1$ it is not still hermitian\,! In order to solve this problem and obtain an hermitian operator for both values of $\sigma\,(=0,1)$\,, we use the change of variable
\be
\w^\a = -\,i\,u^\a ~\sqrt{2N+D-2\,}\,,
\label{c o v}
\ee
where $u^\a$ is a new D-dimensional vector and $N=u\cdot \du$\,. In terms of the new vector, the operator ${\mathbf{K}}_\sigma$ \eqref{K operator} becomes
\bea
\widehat{\mathbf{K}}_\sigma &=& \le(1-\,\tfrac{1}{4}\, u^2\,(\du\cdot\du) \ri) \Box_x
-
\le( u\cdot\p_x-\tfrac{1}{2}\,u^2(\du\cdot\p_x)\ri)\le(\du\cdot\p_x-\tfrac{1}{2}\,u\cdot\p_x(\du\cdot\du)\ri) \nonumber\\
&-&\s\,\m\,\le[~u\cdot\p_x\,-\,u^2\,(\du\cdot\p_x)\,+\,\frac{1}{4}\,u^2\,(u\cdot\p_x)\,(\du\cdot\du)~\ri]~ \frac{1}{\sqrt{2N+D-2}} \nonumber\\
&+&\s\,\m~\frac{1}{\sqrt{2N+D-2}} ~\le[~\du\cdot\p_x\,-\,(u\cdot\p_x)\,(\du\cdot\du)\,+\,\frac{1}{4}\,u^2\,(\du\cdot\p_x)(\du\cdot\du)~\ri] \nonumber\\
&+&\s\,\m^2~\frac{1}{2\,\sqrt{(2N+D)(2N+D-2)}} ~(\du\cdot\du) \,+ \, \s\,\m^2\,u^2 ~\frac{1}{2\,\sqrt{(2N+D)(2N+D-2)}} \nonumber\\
&+&\s\,\m^2~\frac{1}{{2N+D-2}}~\le[~1\,+\,\frac{1}{4}\, u^2\,(\du\cdot\du) ~\ri]
\,, \label{kinetic}
\eea
which is obviously hermitian for any $\s$\,. In a more compact and nicer form, the kinetic operator (Einstein-like operator) \eqref{kinetic} can be written as
\be
\widehat{\mathbf{K}}_\sigma = \le(1-\,\frac{1}{4}\, u^2\,(\du\cdot\du) \ri) \Box_x \,-\,\mathbf{L} \overline{\mathbf{L}}\,,
\label{kinetic 1}
\ee
where
\bea
\mathbf{L}&=&  u\cdot\p_x\,-\,\frac{1}{\,2\,}~u^2\,(\du\cdot\p_x)\, -\, \sigma\, \m \le( a~ \Pi \,+\,\tfrac{1}{\,2\,}~u^2~b \ri)\,, \label{L}\\
\overline{\mathbf{L}}&=& \du\cdot\p_x\,-\,\frac{1}{\,2\,}~(u\cdot\p_x)(\du\cdot\du)\,+\,\sigma \,\m \le( a~ \Pi \,+\,\tfrac{1}{\,2\,}\,b~(\du\cdot\du)\ri)\,,\label{L bar}
\eea
and the quantities of $a$\,, $b$ and $\Pi$ are defined as
\be
a\,=\,\frac{1}{\,\sqrt{2N+D-2}\,} \,,\quad\quad\quad\quad \quad b\,=\,\frac{1}{\,\sqrt{2N+D}\,} \,, \label{a,b}
\ee
\be
\Pi\,=\,1\,-\, u^2~ \frac{1}{\,2(2N+D)\,}~(\du\cdot\du)\,. \label{piii}
\ee
Notice that in above $a$, $b$, $\Pi$ and $(\mathbf{L}\overline{\mathbf{L}})$ are hermitian operators and $(\mathbf{L})^\dag=-\,\overline{\mathbf{L}}$\,. Notice also that, by introducing an alternative change of variable, by considering a positive coefficient in \eqref{c o v}\,, there is another possibility to write an alternative kinetic operator, by substituting $\m \rightarrow -\,\m$\, in \eqref{L}\,, \eqref{L bar}\,.
We mention the obtained kinetic operator in above is indeed the one proposed by Metsaev at the flat spacetime limit \cite{Metsaev: B CSP}
\footnote{\,Note that, the proposed action in \cite{Metsaev: B CSP}\, is written through the use of creation $\a^\m, v$ and annihilation $\bar{\a}^\m, \bar{v}$ operators instead of the auxiliary vector $u^\m$\,. Therefore, at the flat spacetime limit, the kinetic operator of \cite{Metsaev: B CSP} will be coincided with \eqref{kinetic 1} by change of variable $\a^\m = \frac{\sqrt{N_\a}}{v}\, u^\m$\,. }. The use of operators $\mathbf{L}$\, and $\overline{\mathbf{L}}$\,, which actually simplify the form of the kinetic operator, were first discovered in \cite{Metsaev:2008fs}\,. We will obtain these continuous spin operators in the section \ref{LLbar}\,, from the infinite spin limit of the massive higher-spin ones\,.

Finally, using the hermitian kinetic operator \eqref{kinetic 1}\,, we would be able to write the modified bosonic CSP action (when $\s=1$)\,, or the Fronsdal action (when $\s=0$)\,, as
\be
\mathbf{S}_\sigma\,=\,\frac{1
}{\,2\,}~\int \, d^{D}x ~ \mathbf{\Phi}(x,\du) ~\widehat{\mathbf{K}}_\sigma ~\mathbf{\Phi}(x,u) ~\bigg|_{u=0}\,,  \label{action}
\ee
where the gauge field $\mathbf{\Phi}(x,u)$ is now double traceless $({\du\cdot\du})^2\,\mathbf{\Phi}(x,u)=0$\,
\footnote{\,In comparison with \cite{ST PRD}\,, remember that the Schuster-Toro action includes the unconstrained gauge field and parameter\,.}.

\subsection{Gauge symmetry}

To obtain the gauge symmetry of the modified CSP action \eqref{action}, we first present the one for the Fronsdal-like equation and then we rewrite it in terms of the modified gauge field and parameter\,. We will follow the similar method in the previous subsection which led to the action\,.

The Fronsdal-like equation \eqref{eom tilde} is invariant under the gauge transformation
\be
\delta \, \widetilde{\Phi} (x,\omega)\,=\,\left(\,\omega \cdot \p_x + i\, \sigma \m \, \right)\, \widetilde{\varepsilon}\, (x,\omega)\,,   \label{Gauge T}
\ee
where the gauge field $\widetilde{\Phi}$ was given in \eqref{G Field omega}\, and the gauge transformation parameter $\widetilde{\varepsilon}$\, is defined as the generating function
\be
\widetilde{\varepsilon}\, (x,\omega)=\sum\limits_{s=1}^\infty ~\varepsilon_s\, (x,\omega)
=\sum\limits_{s=1}^\infty\tfrac{1}{(s-1)!} \, \, \omega^{\m_{1}}   \ldots \omega^{\m_{s-1}} \, \,\varepsilon_{\m_{1}\ldots\m_{s-1}}(x)\,.
\ee
In above, the higher-spin gauge parameters $\varepsilon_s$\, are totally symmetric tensor fields with the trace constrains $\left(\, \p_\omega \cdot \p_\omega \,\right) \, {\varepsilon}_s\, = 0$\,, while the CSP gauge parameter $\widetilde{\varepsilon}$ satisfies the trace-like constraint
\be
\left(\, \p_\omega \cdot \p_\omega \, + \,\sigma\, \right) \, \widetilde{\varepsilon}\, (x,\omega) = 0\,.   \label{epsilon constraint}
\ee
Note again that when $\s=0$\,, the higher-spin gauge fields ${\Phi}_s$\, and parameters $\varepsilon_s$\, should be substituted in \eqref{Gauge T}\, instead of $\widetilde{\Phi}$\, and $\widetilde{\varepsilon}$\,.

Let us now convert the above trace-like constraint to the trace one (similar to the Fronsdal case)\,. For this purpose, we find that \eqref{epsilon constraint} can be reduced (see the appendix \ref{Trans. oper.}) to the trace constraint on the redefined gauge transformation parameter
$
\left(\, \p_\omega \cdot \p_\omega \, \right) \, {\varepsilon}\, (x,\omega) = 0
$\,, by introducing
\be
\widetilde{\varepsilon}\,(x,\omega) = \mathbf{P}_\varepsilon ~ \varepsilon(x,\omega)\,,
\label{redif eps text}
\ee
where $\mathbf{P}_\varepsilon$ is presented in \eqref{P_eps}\,.
At this moment, we are able to rewrite the gauge transformation \eqref{Gauge T} in terms of the redefined gauge field and parameter\,.
Putting \eqref{tildePhi - Phi} and \eqref{redif eps text}\,in \eqref{Gauge T}\,, we will find the gauge transformation as
\be
\delta\, {\mathbf{\Phi}} (x,\omega)\,=\,\left(\omega \cdot \p_x + i\, \sigma \m +  i\, \sigma \m ~  \w^2 ~ \frac{1}{\,(2N+D)(2N+D-2)\,} \right) {\varepsilon}\, (x,\omega) \,,  \label{Gauge T omega}
\ee
where ${\mathbf{\Phi}}$ and ${\varepsilon}$\, are the new gauge field and parameter\,. To obtain the later, we used the relations \eqref{Peps - Pphi}\,, \eqref{Peps - Pphi 1}\,. We can then apply the change of variable which was used in obtaining the hermitian kinetic operator\,.
Therefore, by applying \eqref{c o v}\, and redefining the gauge parameter (${\boldsymbol{\epsilon}}=-\,i\,\sqrt{2N+D-2}~\varepsilon$), the relation \eqref{Gauge T omega} would be
\be
\delta\, {\mathbf{\Phi }} (x,u)\,=\,\left(u \cdot \p_x \,-\,\s \, \m \, \frac{1}{\sqrt{2N+D-2}} \,+\,\s \, \m ~ u^2\,\frac{1}{\sqrt{2N+D}\,{(2N+D-2)}}\, \right) {\boldsymbol{\epsilon}}\, (x,u) \,,  \label{Gauge T u}
\ee
which is the gauge symmetry of the action \eqref{action}\,.
We mention again that the conditions on the modified CSP gauge field and parameter appear now as $({\du\cdot\du})^2\,\mathbf{\Phi}(x,u)=0$\, and $({\du\cdot\du})\,{\boldsymbol{\epsilon}}(x,u)=0$\,.

\subsection{Gauge fixing}

The Wigner equations are a consequence of the equation of motion after gauge fixing \cite{BM}\,. This fact will examine for the modified bosonic Wigner equations in this subsection, and for fermions in \ref{GF Fff}\,.

By varying the action \eqref{action} with respect to the gauge field $\mathbf{\Phi }$\,, and performing a Fourier transformation in $x^\m$\,, we will obtain the equation of motion in the momentum space
\be
\le[~\le(1-\,\tfrac{1}{4}\, u^2\,(\du\cdot\du) \ri) p^2 \,+\,\mathbb{L} \,\overline{\mathbb{L}} ~\ri]{\Phi}(p,u)=0\,,
\ee
where $\mathbb{L} \,\overline{\mathbb{L}}$ and $\Phi$ are the Fourier transforms of $\mathbf{L} \,\overline{\mathbf{L}}$\, and $\mathbf{\Phi}$\,.
If we then choose the modified de Donder gauge condition
\be
\overline{\mathbb{L}}\,{{\Phi }}= \le(p\cdot\du\,-\,\frac{1}{\,2\,}~(p\cdot u)(\du\cdot\du)\,-\,i\,\sigma \,\m \le( a~ \Pi \,+\,\tfrac{1}{\,2\,}\,b~(\du\cdot\du)\ri)\ri)
{{\Phi }}=0\,,
\label{MdD gauge22}
\ee
proposed by Metsaev in \cite{Metsaev: B CSP}\,, the equation of motion simplifies to
\be
p^2~{{\Phi }}(p,u)=0\,. \label{GF 4}
\ee
The residual gauge invariance allows us to impose the tracelessness of ${{\Phi }}$
\be
(\du\cdot\du)\,{{\Phi }}(p,u)=0\,, \label{GF 2}
\ee
and then due to this condition, the modified de Donder gauge condition \eqref{MdD gauge22} becomes
\be
\le(\,p \cdot \du \,-\,i\,\s\,\m~\frac{1}{\,\sqrt{2N+D-2}\,}\, \ri){{\Phi }}(p,u)=0\,.
\label{GF 3}
\ee
Eventually, by defining a gauge-invariant field (similar to the one in \cite{BM})
\be
{{\varphi}}(p,u) = \delta\le(p \cdot u \,+\,i\,\s \, \m \, \tfrac{1}{\sqrt{2N+D-2}} \,-\,i\,\s \, \m ~ u^2\,\tfrac{1}{\sqrt{2N+D}\,{(2N+D-2)}}\,\ri)\,{{\Phi }}(p,u)\,,
\label{GF 1}
\ee
we can leave the space of polynomials in the auxiliary vector $u$ and reformulate the equations \eqref{GF 4}-\eqref{GF 1} in terms of the gauge-invariant distribution
\bea
\le(p \cdot u \,+\,i\,\s \, \m \, \frac{1}{\sqrt{2N+D-2}} \,-\,i\,\s \, \m ~ u^2\,\frac{1}{\sqrt{2N+D}\,{(2N+D-2)}}\, \ri){{\varphi}}(p,u)=0\,,~~&&\label{cspin11x}\\
\bigg({\du \cdot \du}\bigg)\,{{\varphi}}(p,u)=0\,,~~\label{cspin22x}&&\\
\left(\,p \cdot \du \,-\,i\,\s\,\m~\frac{1}{\,\sqrt{2N+D-2}\,}\,\right){{\varphi}}(p,u)=0\,,~~\label{cspin33x}&&\\
p^2~{{\varphi}}(p,u)=0\,.~~&& \label{cspin44x}
\eea
Indeed, the gauge field $\Phi(p,u)$ is a polynomial in the auxiliary vector $u$ while the gauge-invariant field $\varphi(p,u)$ in above equations is a distribution of the form \eqref{GF 1}\,. If we now apply the change of variable \eqref{c o v}\, in the above equations, we will arrive at the modified Wigner equations \eqref{cspin11}-\eqref{cspin44}\,, proposed in the previous section\,.

\subsection{Relation to the Schuster-Toro action }\label{S-T points}

This subsection clarifies a close relationship between the obtained generating functions of \cite{ST PRD} and the introduced operators of this paper ($\mathbf{P}_\Phi$ and $\mathbf{P}_\varepsilon$), used for rearranging of traces, as well as a relationship between the constrained action formalisms of \cite{ST PRD} and \cite{Metsaev: B CSP}\,.

Using the Bessel function of the first kind \eqref{Bessel} and rising Pochhammer symbol \eqref{Pochhammer}, we can write
\be
\sum_{n=0}^{\infty}\, x^{\,2n}\, \frac{(-1)^n}{2^{2n}\, n! \, (\a+1)_n}  = \Gamma(\a+1)\, \le(\tfrac{x}{2}\ri)^{-\a}\, J_\a (x) \,. \label{poch bessel}
\ee
Then, putting $x=\sqrt{-\,\w^2}$, this relation can be rewritten as
\be
\sum_{n=0}^{\infty}\, \w^{\,2n}\, \frac{1}{2^{2n}\, n! \, (N+\tfrac{D}{2})_n}  = \Gamma (N+\tfrac{D}{2})\, \le(\tfrac{x}{2}\ri)^{-N-\frac{D-2}{2}}\, J_{N+\frac{D-2}{2}} (x) \, \bigg|_{x=\sqrt{-\,\w^2}} \,, \label{poch bessel 1}
\ee
for $\a=N+\tfrac{D-2}{2}$, and
\be
\sum_{n=0}^{\infty}\, \w^{\,2n}\, \frac{1}{2^{2n}\, n! \, (N+\tfrac{D}{2}-1)_n}  = \Gamma(N+\tfrac{D}{2}-1)\, \le(\tfrac{x}{2}\ri)^{-N-\frac{D-4}{2}}\, J_{N+\frac{D-4}{2}} (x) \, \bigg|_{x=\sqrt{-\,\w^2}} \,, \label{poch bessel 11}
\ee
for ${\a=N+\tfrac{D-4}{2}}$\,. In these relations, if we then consider $w$ as a D-dimensional vector and $N=\w \cdot \dw$\,, we will immediately find that the left-hand-side of \eqref{poch bessel 1}, \eqref{poch bessel 11} are respectively equal to the operators of $\mathbf{P}_\varepsilon$ and $\mathbf{P}_\Phi$, introduced in \eqref{P_eps}, \eqref{P_phi} (with $\s=1$)\,\footnote{\,Note that the operators in \eqref{P_eps}, \eqref{P_phi} are written by considering the mostly plus signature for the metric while here they are taken into account in the mostly minus signature to be easily compared with results of \cite{ST PRD}\,. }. Recall that these operators were responsible to convert the (double) trace-like constraints to the (double) trace ones. On the other side, looking at the right-hand-side of \eqref{poch bessel 1}, \eqref{poch bessel 11} with $N=0$, we surprisingly discover that they are exactly (up to a coefficient for \eqref{poch bessel 11}) equivalent to the introduced generating functions of \cite{ST PRD}, called by $G(\w)$ and $G'(\w)$ recpectively\,. Therefore, the introduced operators of this paper ($\mathbf{P}_\varepsilon$, $\mathbf{P}_\Phi$), used for obtaining a constrained formalism of the bosonic CSP action, have a similar role as $G(\w)$ and $G'(\w)$ in \cite{ST PRD}, applied to get a constrained form of the action (but after a partial gauge fixing)\,.

To make clear the relationship between the tensor form of the action in \cite{ST PRD} and the Metsaev action \cite{Metsaev: B CSP} in flat spacetime, we consider \eqref{action} with the mostly negative signature for the metric which, using a change of auxiliary variable
\be
\sqrt{2}\,\e^\a = -\,u^\a\,\sqrt{2N+{D}-2}\,,
\ee
becomes
\be
\mathbf{S}_\sigma\,=\,-\,\frac{1
}{\,2\,}~\int \, d^{D}x ~ \mathbf{\Phi}(x,\p_\e) ~\widehat{\mathcal{K}}_\sigma ~\mathbf{\Phi}(x,\e) ~\bigg|_{\e=0}\,,  \label{action eta}
\ee
where
\be
\widehat{\mathcal{K}}_\sigma = \le(1-\,\frac{1}{4}\, \e^2\,(\p_\e\cdot\p_\e) \ri) \Box_x \,+\,\mathcal{L} \overline{\mathcal{L}}\,,
\label{kinetic 1 ST}
\ee
\be
\overline{\mathcal{L}}= \p_\e \cdot\p_x - \frac{1}{\,2\,}~(\e\cdot\p_x)(\p_\e \cdot\p_\e) + \tfrac{\s\m}{\sqrt{2}(N+\frac{D}{2} -1)}\le(1-\e^2\,\tfrac{1}{2(2N+D)}\,\p_\e \cdot\p_\e\ri) -\tfrac{\s\m}{2\sqrt{2}}\,\p_\e \cdot\p_\e
\,,\label{L bar eta}
\ee
and $(\mathcal{L})^\dag=-\,\overline{\mathcal{L}}$\,.
Then, using the generating function of the modified CSP gauge field
\be
\mathbf{\Phi}(x,\e)\,=\,\sum_{s=0}^{\infty}~\,\frac{1}{\,s!\,} ~ \e^{\m_{1}}   \ldots \e^{\m_{s}} \, \,
\overset{(s)}{\Phi}_{\m_{1}\ldots\m_{s}}(x) \,,
\label{G Field eta}
\ee
we can compute, for instance, the second term of the action \eqref{action eta} as
\bea
\mkern-30mu\mathbf{S}_{II}&=& \frac{1}{8}\,\int d^{D}x ~ \mathbf{\Phi}(x,\p_\e) \le[\e^2\,(\p_\e\cdot\p_\e) \,\Box_x \ri] \mathbf{\Phi}(x,\e) ~\bigg|_{\e=0}\,,\\
\mkern-30mu&=&  \frac{1}{8}\,\int d^{D}x ~\sum_{s=0}^{\infty}\,\frac{s(s-1)}{s!\,s!}\, \overset{(s)}{\Phi}_{\n_{1}\ldots\n_{s}} \,
\le( \p_\e^{\n_1} \ldots \p_\e^{\n_s} \, \e^{\m_{1}}   \ldots \e^{\m_{s}} \ri)(2 \,g_{\m_1\m_2})\,\Box_x\, \overset{(s)}{\Phi\, '}_{\m_{3}\ldots\m_{s}}\,,\\
\mkern-30mu&=&-\,\frac{1}{4}\,\int d^{D}x ~\sum_{s=0}^{\infty}\,\frac{s(s-1)}{s!}\, \bigg( \p^\a \, \overset{(s)}{\Phi \, '}{}^{\m_{1}\ldots\m_{s-2}}\bigg)~
\bigg( \p_\a\,\overset{(s)}{\Phi \,'}_{\m_{1}\ldots\m_{s-2}} \bigg)\,,
 \eea
where $\Phi \,'$ stands for the trace of the gauge field $\Phi$\,. The final result for all terms of the action \eqref{action eta} yields
\be
\mathbf{S}_\sigma=\sum_{s=0}^{\infty} \frac{1}{s!} \int d^D x \le[~ \frac{1}{2}\,\Big(\partial_\mu \Phi^{(s)}\Big)^2 \,-\, \frac{s(s-1)}{4}\, \Big(\partial_\mu {\Phi^{(s)}}' \Big)^2 \,-\, \frac{s}{2}\,\Big(\mathbf{D}_\s^{(s-1)}\Big)^2 ~\ri] \label{action rank}
\ee
where
\be
\mathbf{D}_\s^{(s-1)} \equiv \partial_x\cdot \Phi^{(s)} - \tfrac{1}{2}\partial_x \circ {\Phi^{(s)}}'+  \tfrac{\s\m}{\sqrt{2} (s+\frac{D}{2}-2)} \left(\Phi^{(s-1)}- \tfrac{1}{ (2s+D-6)}\, g\circ {\Phi^{(s-1)}}' \right) - \tfrac{\s\m}{2 \sqrt{2}}\,{\Phi^{(s+1)}}'\,, \label{DDD}
\ee
and the notations ``$.$'' and ``$\circ$'' were used, for simplicity, from \cite{ST PRD}\,. The action \eqref{action rank} is written in tensor form such that when $\s=0$ the D-dimensional Schwinger-Fronsdal action, which is rank-diagonal, can be read\,. When $\s=1$\,, the action is rank-mixing and describes the constrained formalism of a single CSP in D-dimensions. For $D=4$, the action \eqref{action rank} was first found by Schuster and Toro \cite{ST PRD} which is precisely the Metsaev action \cite{Metsaev: B CSP} in flat-space\,.



\section{Modified fermionic continuous spin gauge field}  \label{MFCSP}

The modified fermionic CSP action in (A)dS spacetime was recently presented in \cite{Metsaev: F CSP}\,.
The action is written in a fashion similar to the Fang-Fronsdal action which includes a triple gamma-trace constraint on the gauge field and a gamma-trace constraint on the gauge parameter\,.
In \cite{BMN} it is shown that, by performing a Fourier transform in the auxiliary vector, the Fang-Fronsdal-like equation will lead to the fermionic CSP action, presented in \cite{BNS}\,. However, the relation of the Fang-Fronsdal-like equation and the modified fermionic CSP action has not been examined\,. This section will make clear this relation by performing a field redefinition in the Fang-Fronsdal-like equation\,.
The modified gauge symmetry and modified Wigner equations, as a consequence of the equation of motion, will be obtained as well\,.

\subsection{CSP action}

We introduce the fermionic continuous spin gauge field as
\be
\widetilde{\Psi}(x,\omega)\,=\,\sum_{n=0}^{\infty}~{\Psi}_n(x,\omega)\,=\,\sum_{n=0}^{\infty}\,\frac{1}{\,n!\,} ~ \omega^{\m_{1}}   \ldots \omega^{\m_{n}} \, \,
{\Psi}_{\m_{1}\ldots\m_{n}}(x)\,,  \label{generating func.}
\ee
where ${\Psi}_n$ denote for all totally massless symmetric spinor-tensor fields of spin $s=n+\frac{1}{2}$\,, and $\w^\m$ is a D-dimensional auxiliary vector\,. Then, the Fang-Fronsdal and Fang-Fronsdal-like equations can be written as \cite{BM}
\be
\widetilde{\mathbb{F}}_\sigma~\widetilde{\Psi}(x,\omega)=\Big[\,\g \cdot \p_x - \left(\omega \cdot \p_x + i\, \sigma \m\right) \left(\g \cdot \p_{\omega} + \,i\, \s\, \right)\,\Big]\,\widetilde{\Psi}(x,\omega) =0\,, \label{eom tilde F}
\ee
where $\s$ was introduced in \eqref{sigma}\,. The triple gamma-trace-like constraint on the gauge field is given by
\be
\le(\g \cdot \dw + i\,\s \ri)\le(\dw \cdot \dw +\,\s \ri)\widetilde{\Psi}(x,\omega) =0\,. \label{Field cons. F}
\ee
By referring to the appendix \ref{Trans. oper.}\,, the above condition can be reduced to the triple gamma-trace constraint on the modified gauge field
$
\le(\g \cdot \dw \ri)^3 {\mathbf{\Psi}}(x,\omega) =0 \label{Field cons. F2}
$\,, by introducing
\be
\widetilde{\Psi}(x,\omega) = {\mathbf{P}}_\Psi ~ {\mathbf{\Psi}}(x,\omega)\,,
\label{Psitild to Psi}
\ee
where
\be
\mathbf{P}_\Psi =\sum_{k=0}^{\infty}\le[(\sigma\,\gamma \cdot \omega)^{2k} + 2ik\, (\sigma\,\gamma \cdot \omega)^{2k-1}\ri]
\frac{(-1)^k}{~2^{\,2k}~k!~(N+\tfrac{D}{2}-1)_k~}\,.
\ee
Now we rewrite the Fang-Fronsdal-like equation in terms of the modified gauge field\,. To this end, we substitute \eqref{Psitild to Psi} in \eqref{eom tilde F}\,, and then using \eqref{1 1}-\eqref{3 3}\,, the equation \eqref{eom tilde F} becomes
\be
{\mathbb{F}}_\sigma~{\mathbf{\Psi}}(x,\omega)=0\,,
\ee
where
\bea
\mkern-40mu {\mathbb{F}}_\sigma&=&\,\g \cdot \p_x \, -\, \left(\omega \cdot \p_x \right) \left(\g \cdot \p_{\omega}  \right) \,-\,\s\, \m \,\frac{2}{2N+D-2} \, -\, i\,\s\,\m \left(\g \cdot \p_{\omega}  \right)   \nonumber \\
&&~\quad\quad\,\,+\,  i\,\s\,\m \left(\g \cdot \w  \right) \, \frac{2}{(2N+D-2)^2}
\, -\, i\,\s\,\m\,\w^2\,\frac{1}{(2N+D)^2}\,\left(\g \cdot \p_{\omega}  \right)  \nonumber\\
&&~\quad\quad\,\,-\,\s\,\m\left(\g \cdot \w  \right)\, \frac{2}{(2N+D)(2N+D-2)} \,\left(\g \cdot \p_{\omega}  \right) \,+\, \mathcal{O}(\w^3)
  \,.  \label{eom F}
\eea
Note that the terms comprising the powers of $\w^3$ (or higher)\, will be vanished (at the level of the action)\,, due to the triple gamma-trace constraint ${\overline{\mathbf{\Psi}}}(x,\dw)\,(\g\cdot\w)^3=0$\,, and hence we will not consider them in the rest of calculations\,.

By applying the hermitian conjugates \eqref{hermitian conjugates}\,, we find that the operator ${\mathbb{F}}_\sigma$ is not anti-hermitian $\mathbb{F}_\s^{\,\dag} \neq -\,\g^0\,\mathbb{F}_\s\,\g^0$\,. To obtain an anti-hermitian operator, we should multiply \eqref{eom F} by
\be
1\,-\, \frac{1}{2}\,(\g\cdot\omega)(\g\cdot\p_\omega)\,-\,\frac{1}{4}\,\w^2\,(\dw\cdot\dw)
\ee
to the left, which consequently leads to the operator
\bea
\mkern-40mu\mathbb{K}_\s&=& \,\le(1\,-\,\tfrac{1}{4}\,\w^2\,(\dw\cdot\dw) \ri)\g\cdot\p_x \,-\, (\g\cdot\w)(\dw\cdot\p_x) \,-\, (\w\cdot\p_x)(\g\cdot\dw)   \nonumber \\
&& ~~ +\, \frac{1}{2}\,(\g\cdot\w)(\w\cdot\p_x)(\dw\cdot\dw) \,+\, \frac{1}{2}\, \w^2\, (\dw\cdot\p_x)(\g\cdot\dw)\,+\, (\g\cdot\w)(\g\cdot\p_x)(\g\cdot\dw)   \nonumber \\
&& ~~ -\,\s\,\m\,\Big[\,1\,-\,(\g\cdot\omega)(\g\cdot\p_\omega)\,-\,\frac{1}{4}\,\w^2\,(\dw\cdot\dw)  \Big] ~\frac{2}{2N+D-2}\nonumber \\
&& ~~  -\,i\,\s\,\m\,(\g \cdot \w) ~\frac{1}{2N+D-2} \,+\,i\,\s\,\m\,\frac{1}{2}\,(\g\cdot\w)(\dw\cdot\dw) \nonumber \\
&& ~~\,-\,i\,\s\,\m\,(\g\cdot\dw) \,+\,i\,\s\,\m\,\frac{1}{4}~\w^2~\frac{2}{2N+D}~(\g\cdot\dw)
    \,.  \label{eom F 2}
\eea
This operator is obviously anti-hermitian $\mathbb{K}_\s^{\dag}=-\,\g^0\,\mathbb{K}_\s\,\g^0$\, when $\s=0$\,, however when $\s=1$ it is not anti-hermitian\,.
Similar to the bosonic case, if we apply the change of variable \eqref{c o v} in \eqref{eom F 2}\,, we will find then an anti-hermitian operator for the both values of $\s$\,, which is\,\footnote{\,The operator $\widehat{\mathbb{K}}_\s$ is an anti-hermitian operator while $-\,i\,\widehat{\mathbb{K}}_\s$\,, which recall it as the fermionic kinetic operator, is hermitian\,.}
\bea
\mkern-40mu \widehat{\mathbb{K}}_\s&=& \, \,\le(1\,-\,\tfrac{1}{4}\,u^2\,(\du\cdot\du) \ri)\g\cdot\p_x \,-\, (\g\cdot u)(\du\cdot\p_x) \,-\, (u\cdot\p_x)(\g\cdot\du)   \nonumber \\
&& ~~ +\, \frac{1}{2}\,(\g\cdot u) (u\cdot\p_x)(\du\cdot\du) \,+\, \frac{1}{2}\, u^2\, (\du\cdot\p_x)(\g\cdot\du)\,+\, (\g\cdot u)(\g\cdot\p_x)(\g\cdot\du)  \nonumber \\
&& ~~-\,\s\,\m\,\le[\,1\,-\,(\g\cdot u)(\g\cdot\p_u)\,-\,\frac{1}{4}\,u^2\,(\du\cdot\du)  \,\ri] ~\frac{2}{2N+D-2}\nonumber \\
&& ~~-\,\s\,\m\,\le[\,\g \cdot u \,-\,\frac{1}{2}\,u^2\,(\g\cdot \p_u)\,\ri]~\frac{1}{\sqrt{2N+D-2}} \nonumber \\
&& ~~+\,\s\,\m~\frac{1}{\sqrt{2N+D-2}}~\le[\,\g \cdot \p_u \,-\,\frac{1}{2}\,(\g\cdot u)(\p_u\cdot\p_u)\,\ri]
\,.  \label{eom F 3}
\eea
This operator is precisely the flat spacetime limit of the one proposed in (A)dS space by Metsaev \cite{Metsaev: F CSP}
\footnote{\,Note that, similar to the bosonic action, the proposed action in \cite{Metsaev: F CSP}\, is written through the use of creation $\a^\m, v$ and annihilation $\bar{\a}^\m, \bar{v}$ operators instead of the auxiliary vector $u^\m$\,. Thus, at the flat spacetime limit, the kinetic operator of \cite{Metsaev: F CSP} will be coincided with \eqref{eom F 3} by change of variable $\a^\m = \frac{\sqrt{N_\a}}{v}\, u^\m$\,. }. Note that, similar to the bosonic case, the alternative change of variable will lead to another operator by replacing $\m \rightarrow -\,\m$\, in \eqref{eom F 3}\,.

Using the above operator, we can then write the modified fermionic CSP action (when $\s=1$)\,, or the Fang-Fronsdal action (when $\s=0$)\,, as
\be
\mathbf{S}_\s =\,-\, i \int d^D x ~ \overline{\mathbf{\Psi}}(x,\p_u)
~\widehat{\mathbb{K}}_\s ~{\mathbf{\Psi}}(x,u) ~\Big|_{u=0}\,,
\label{action F}
\ee
where $\overline{\mathbf{\Psi}}={\mathbf{\Psi}}^{\,\dag}\,\g^0$ is the Dirac adjoint and the gauge field ${\mathbf{\Psi}}$ is triple gamma-traceless $\le(\g \cdot \du \ri)^3 {\mathbf{\Psi}}(x,u) =0$\,\footnote{\,Remind that the proposed action in \cite{BNS} includes the unconstrained gauge field and parameter\,.}.

\subsection{Gauge symmetry}

By presenting the gauge symmetry of the Fang-Fronsdal-like equation, we will rewrite it in terms of the redefined gauge field and parameter to obtain the one for the modified CSP action\,.

The gauge symmetry of the Fang-Fronsdal-like equation \eqref{eom tilde F} is given by
\be
\delta \, \widetilde{\Psi} (x,\omega)\,=\,\left(\omega \cdot \p_x + i\, \sigma \m \right)\, \widetilde{\xi}\, (x,\omega)\,,   \label{Gauge T F}
\ee
where the spinor gauge parameter
\be
\widetilde{\xi}\, (x,\omega)=\sum\limits_{n=1}^\infty ~\xi_n\, (x,\omega)
=\sum\limits_{n=1}^\infty\tfrac{1}{(n-1)!} \, \, \omega^{\m_{1}}   \ldots \omega^{\m_{n-1}} \, \,\xi_{\m_{1}\ldots\m_{n-1}}(x)\,,
\ee
is gamma traceless-like
$
\left(\, \g \cdot \p_\omega \,+\,i\, \sigma\, \right) \, \widetilde{\xi}\, (x,\omega) = 0
$\,.
On the other hand, a gamma trace constraint
$
\le(\,\g \cdot \dw \,\ri)\,\xi(x,\w)=0
$\,, can be constructed by (see the appendix \ref{Trans. oper.})
\be
\widetilde{\xi}(x,\omega) =\mathbf{P}_\xi ~\xi(x,\omega)\,,
\label{xi tlid xi tex}
\ee
where $\xi$ is a redefined gauge parameter and the operator $\mathbf{P}_\xi $ is presented in \eqref{P xi}\,.
Now we can rewrite the gauge symmetry using the new gauge field and parameter\,. To do that, we apply \eqref{Psitild to Psi}\, and \eqref{xi tlid xi tex}\, in the gauge transformation \eqref{Gauge T F}\,, which yields
\be
\mkern-11mu\delta\,{\mathbf{\Psi}}=\le[\w\cdot \p_x +\,i\s\m \,-\, \s\m\,(\g\cdot\w)\,\frac{2}{(2N+D)(2N+D-2)}\,+\,i\s\m\,\w^2\,\frac{1}{(2N+D)^2} \,\ri]\xi\,.
\label{Gauge T F 1}
\ee
To obtain the later, the relations \eqref{Pkesi--Ppsi}\, and \eqref{Pkesi--Ppsi 1}\, are used\,.
Since the action was written in terms of $u$-vector, we employ the change of variable \eqref{c o v}\, and a redefined gauge parameter ($\zeta=-\,i\,\sqrt{2N+D-2}~\xi$) in \eqref{Gauge T F 1}\, to obtain
\be
\mkern-11mu\delta\,{\mathbf{\Psi}}=\le[\,u\cdot \p_x \,-\,\s\m\,\tfrac{1}{\sqrt{2N+D-2}} \,-\, \s\m\,(\g\cdot u)\,\tfrac{2}{(2N+D)(2N+D-2)} \,+\, \s\m\,u^2\,\tfrac{1}{(2N+D)^{3/2}} \,\ri]{\zeta}\,,
\label{Gauge T F 2}
\ee
which is the gauge symmetry of the action \eqref{action F}\,. We remind again that the constraint on the modified CSP gauge field is
$
\le(\g \cdot \du \ri)^3 {\mathbf{\Psi}}(x,u) =0 \label{Field cons. F2 u}
$\, and the one for the gauge parameter is
$
\le(\,\g \cdot \du \,\ri)\,\zeta(x,u)=0
$\,.

\subsection{Gauge fixing} \label{GF Fff}

Varying the action \eqref{action F} with respect to the gauge field $\mathbf{\Psi}$\,, we get the equation of motion \,$
\widehat{\mathbb{K}}_\s ~{\mathbf{\Psi}}(x,u)=0
$\,, which after dropping the multiplier ${ 1\,-\, \frac{1}{2}\,(\g\cdot u)(\g\cdot\du)\,-\,\frac{1}{4}\,u^2\,({\du \cdot \du}) }$\,, and performing a Fourier transform in $x^\m$, would be corresponding to
\bea
\mkern-13mu&&\bigg[\,\g \cdot p \, -\, \left(p \cdot u \right) \left(\g \cdot \du
\right) \,+\,i\,\s \,\m \,\frac{2}{2N+D-2}  \label{eom ricci} \\
\mkern-13mu&&\,-\,i\,\s\,\m \left(\g \cdot u  \right) \, \frac{2}{(2N+D-2)^{3/2}}
\,-\,i\,\s\,\m\,\frac{1}{\sqrt{2N+D-2}}\, \left(\g \cdot \du \right) \nonumber\\
\mkern-13mu&&\,+\,i\s\m\left(\g \cdot u  \right)\, \frac{2}{(2N+D)(2N+D-2)} \,\left(\g \cdot \du  \right)+i\s\m u^2\,\frac{1}{(2N+D)^{\,3/2}}\,\left(\g \cdot \du  \right)\bigg]
{{\Psi(p,u)}}=0
  \,. \nonumber
\eea
If we now choose the gauge
\be
{\cal P}_1\,{{\Psi}} = \left(\g \cdot \du  \right){{\Psi}}=0\,,
\label{GF I}
\ee
the equation of motion \eqref{eom ricci} reduces to
\be
{\cal P}_2\,{{\Psi}} = \le(\,\g \cdot p \,+\,i\,\s \,\m \,\frac{2}{2N+D-2}
\,-\,i\,\s\,\m \left(\g \cdot u  \right) \,  \, \frac{2}{(2N+D-2)^{3/2}} \,\ri){{\Psi}}=0\,,
\label{GF II}
\ee
where ${\cal P}_1$\,, ${\cal P}_2$ are two operators\,. Afterwards, using the anti-commutator of the introduced operators, we can illustrate that
\be
\le\{\,{\cal P}_1 \,,\, {\cal P}_2\, \ri\}\,{{\Psi}} = \,2 \,\le(\,p\cdot \du \,-\,i\,\s\,\m\, \frac{{2N+D}}{(2N+D-2)^{3/2}} \, \ri)\,{{\Psi}}=0\,.
\label{GF III}
\ee
Then, similar to the bosonic case, if we introduce a gauge invariant field
\be
\mkern-16mu{{\psi}} =\delta \le(\,p \cdot u \,+\,i\s\m\,\tfrac{1}{\sqrt{2N+D-2}} \,+\, i\s\m\,(\g\cdot u)\,\tfrac{2}{(2N+D)(2N+D-2)} \,-\, i\s\m\,u^2\,\tfrac{1}{(2N+D)^{3/2}} \, \ri)  {{\Psi}}\,,
\label{GF IV}
\ee
we will be able to write the relations \eqref{GF I}-\eqref{GF IV} as the following
\bea
\le(\,p \cdot u \,+\,i\s\m\,\tfrac{1}{\sqrt{2N+D-2}} \,+\, i\s\m\,(\g\cdot u)\,\tfrac{2}{(2N+D)(2N+D-2)} \,-\, i\s\m\,u^2\,\tfrac{1}{(2N+D)^{3/2}} \, \ri){{\psi}}=0\,,~~&&\label{fcspin11g}\\
\Big(\g\cdot\du \Big)\,{{\psi}}=0\,,~~\label{fcspin22g}&&\\
\left(\,p\cdot \du \,-\,i\,\s\,\m\, \tfrac{{2N+D}}{(2N+D-2)^{3/2}} \, \right){{\psi}}=0\,,~~\label{fcspin33g}&&\\
\le(\,\g \cdot p \,+\,i\,\s \,\m \,\tfrac{2}{2N+D-2}
\,-\,i\,\s\,\m \left(\g \cdot u  \right) \,  \, \tfrac{2}{(2N+D-2)^{3/2}} \,\ri)\,{{\psi}}=0\,.~~&& \label{fcspin44g}
\eea
By applying the change of variable \eqref{c o v} in above equations, we will arrive at the modified Wigner equations \eqref{fcspin11}-\eqref{fcspin44}\,, except\,\footnote{\,A similar exception was happened in \cite{BM} when the Wigner equations were obtained from the Fang-Fronsdal-like equation by fixing the gauge\,. } the second one \eqref{fcspin22g} which should be multiplied by
\be
\left(\,{\g \cdot \du} \,-\,\s\,(\g\cdot u)(\g\cdot\du)\,\tfrac{\sqrt{2N+D}}{2N+D+2}\,-\,\s\,\tfrac{2\,\sqrt{2N+D}}{2N+D+2} \,\right)\,.
\ee

\section[Modified CSP action from the massive higher-spin action]{Modified CSP action from the massive higher-spin action\footnote{\,We thank Ruslan Metsaev for introducing the Ref. \cite{Metsaev:2008fs} which rose this section to the draft\,.}}\label{LLbar}


Khan and Ramond discovered that the continuous-spin representation can be obtained from the massive higher-spin representation by taking the limit $m\rightarrow 0$\,, $s \rightarrow \infty$\, and $ms\,=\,\m\,=\,\mbox{constant}$\, \cite{KR}\,.
This issue was utilized in \cite{BM} which led to find the (Fang-) Fronsdal-like equation from the bosonic (fermionic) massive higher-spin equations\,. In this section, by following the same procedure as \cite{BM}\,, we will demonstrate how to obtain the modified bosonic (fermionic) CSP action from the massive bosonic (fermionic) spin-$s$ fields action, by taking the limit\,. Here in this paper, we will examine the cases in flat spacetime, however, a similar fashion can be applied to the cases in (A)dS spacetime\,.

\subsection{Massive bosonic higher-spin action}

Using an auxiliary space, we can introduce the massive bosonic spin-$s$ gauge fields by the generating functions
\be
\varphi_s (x,u,v)\,=\,\sum\limits_{s'=0}^s{\frac{1}{s'!\,(s-s')!}}~\,u^{\mu_1}\dots u^{\mu_{s'}}~v^{s-s'}~\varphi_{\mu_1\dots\,\mu_{s'}}(x)\,,
\label{varphi}
\ee
where $u^\m$ is a D-dimensional auxiliary vector and $v$ is an auxiliary scalar\,. The gauge fields are double traceless $(\du\cdot\du)^2\,\varphi_s=0$ and homogeneous polynomials of degree $s$ in both $u$ and $v$ variables, implying
\be
\le(\,N_u \,+\, N_v \,-\, s \,\ri)\varphi_s=0\,.
\label{homogen}
\ee
We notice here that, the relation \eqref{homogen} is clearly ill defined in the limit $s\rightarrow \infty$\,,
however by introducing the variable $\a=v/s$\, and the new gauge fields $\varphi_s(u,v)=\a^s\,\Phi_s(u,\a)$\,, after dropping the singular factor $\a^s$\,, will remain well defined $\le(N_u + N_\a\ri)\Phi_s=0$\,.

From \cite{Metsaev:2008fs}\,, the gauge invariant action of massive spin-$s$ fields in D-dimensional flat spacetime can be given by
\be
\mathbf{S}_{(m,s)}\,=\,\frac{1
}{\,2\,}~\int \, d^{D}x ~~ {\varphi_s}(x,\du,\p_v) ~\widehat{\mathbf{K}}_{(m,s)} ~{\varphi_s}(x,u,v) ~\bigg|_{u,v=0}\,,  \label{action m}
\ee
where the kinetic operators have a compact form as \be
\widehat{\mathbf{K}}_{(m,s)}  =\le(1-\,\tfrac{1}{4}\, u^2\,(\du\cdot\du) \ri)\Big( \Box_x - m^2 \Big) \,-\,{\mathbf{C}}_{(m,s)} \,\overline{\mathbf{C}}_{(m,s)} \,,
\label{kinetic m}
\ee
and
\bea
{\mathbf{C}}_{(m,s)}&=& u\cdot\p_x\,-\,\frac{1}{\,2\,}~u^2\,(\du\cdot\p_x)\,-\,\frac{1}{2}\,m\,A_s~\p_v\,u^2\,-\,m\,v\,A_s\,\Pi \,,  \\
\overline{\mathbf{C}}_{(m,s)}&=& \du\cdot\p_x\,-\,\frac{1}{\,2\,}~(u\cdot\p_x)(\du\cdot\du)\,+\,\frac{1}{2}\,m\,v\,A_s~(\du\cdot\du)\,+\,m\,{\Pi}\,A_s~\p_v\,,
\label{L bar m}
\eea
\bea
A_s\,&=&\,\le(\frac{2s+D-4-N_v}{2s+D-4-2N_v} \ri)^{\frac{1}{2}} \,,
\qquad\qquad~~\, N_v = v\,\p_v \,,
\\
\Pi\,&=&\,1\,-\, u^2~ \frac{1}{\,2(2N_u+D)\,}~(\du\cdot\du)\,,
\qquad N_u = u\cdot\du\,.
\eea
Note that, the kinetic operators have indeed a complicated form but when are written in terms of the operators ${\mathbf{C}}$ and $\overline{\mathbf{C}}$\,, found first in \cite{Metsaev:2008fs}\,, will take a nice form as above (compare e.g. \eqref{kinetic} and \eqref{kinetic 1} in the CSP case)\,. Varying the action with respect to the fields, we will find the equations of motion
\be
\widehat{\mathbf{K}}_{(m,s)} ~{\varphi_s}(x,u,v)=0\,.
\label{kinetic m 1}
\ee
Now, let us to take the limit of infinite spin from the equations of motion \eqref{kinetic m 1}\,. Then, it would be expected to find the kinetic operator of the modified bosonic CSP action \eqref{kinetic 1}\,, and consequently the action \eqref{action}\,.
From \eqref{kinetic m} and \eqref{kinetic m 1}\,, it is seen that $m^2$ will be vanished in the limit, therefore it would be enough to concentrate our discussion on the operator $\overline{\mathbf{C}}$ through the modified de Donder gauge conditions\,. Recall that the operator $\mathbf{C}$ will be then obtained via $(\overline{\mathbf{C}})^\dag=-\,{\mathbf{C}}$\,, by applying the following introduced hermitian conjugates
\be
(\p_x^{\,\a})^\dag:=\,-\,\p_x^{\,\a} \,,\qquad (\du^{\,\a})^\dag := u^{\,\a} \,,\qquad (v)^\dag:=\p_v \,.
\ee
\vspace{1mm}
\noindent\textbf{Modified de Donder gauge conditions}

\noindent The modified de Donder gauge conditions\footnote{\,It was called ``de Donder-like gauge condition'' in \cite{Metsaev:2008fs}\,.}of the massive spin-$s$ fields, which actually simplify the equations of motion, are given by
\be
\overline{\mathbf{C}}_{(m,s)} ~\varphi_s(x,u,v)=0\,,
\label{MdD}
\ee
where $\varphi_s$ and $\overline{\mathbf{C}}$ introduced in \eqref{varphi} and \eqref{L bar m}\,. Now, if we introduce the variable $\a=v/s$\,, the parameter $\m=ms$\, and
\be
\varphi_s(x,u,v)=\a^s\,\mathbf{\Phi}_s(x,\w) \,,
\qquad \w^{\,\m}={u^{\,\m}}/{\a}\,,
\label{singular f}
\ee
then the modified de Donder gauge conditions \eqref{MdD} become
\bea
\mkern-10mu&&\bigg[\,\dw\cdot\p_x\,-\,\frac{1}{\,2\,}~(w\cdot\p_x)(\dw\cdot\dw)\,+\,\frac{1}{2}~\m\,\le(\frac{N_\w+D-2+s}{2N_\w+D} \ri)^{\frac{1}{2}}(\dw\cdot\dw) \label{MdD omega} \\
\mkern-10mu&& \quad\quad\quad\quad\quad\quad\quad\quad\quad\quad\quad\quad\quad\quad +~\m\,\Pi\,\le(\frac{N_\w+D-3+s}{2N_\w+D-2} \ri)^{\frac{1}{2}} \frac{1}{s^2}\,\big(\,s-N_\w\big)\, \bigg]\,\mathbf{\Phi}_s(x,\w)=0\,, \nonumber
\eea
where $N_\w=\w\cdot\dw$\, and the following useful relations have been used
\be
{\du}=\frac{1}{\a}\,{\p_\omega}\,, \qquad
N_v\,=\,-\,N_\w\,+\,N_\a\,.
\ee
Nevertheless, it is seen that, in spite of dropping the singular factor from \eqref{MdD omega}\,, it is not still well defined in the limit\,. However, we can apply the change of variable
\be
\w^{\a}\,=\,\mathbf{u}^{\,\a}~\frac{1}{\sqrt{{N_\mathbf{u}}+D-2+s}}\,,
\ee
where $\mathbf{u}^{\,\a}$ is a D-dimensional vector and $N_\mathbf{u}=\mathbf{u}\cdot \p_\mathbf{u}$\,, to acquire the well defined modified de Donder gauge conditions, which are
\bea
\mkern-60mu&&\overline{\mathbb{C}}_{(m,s)}~\mathbf{\Phi}_s(x,\mathbf{u})=\bigg[\,\p_{\mathbf{u}}\cdot\p_x\,-\,\frac{1}{\,2\,}~(\mathbf{u}\cdot\p_x)(\p_{\mathbf{u}}\cdot\p_{\mathbf{u}})\,+\,\frac{1}{2}~\m\,
\le(
\frac{1}{\sqrt{2N_\mathbf{u}+D}}\ri)(\p_{\mathbf{u}}\cdot\p_{\mathbf{u}}) \nonumber\\
\mkern-60mu&& \qquad\qquad\qquad\qquad\qquad\qquad
+\,\m\,\Pi\,\le(
\frac{N_\mathbf{u}+D-3+s}{\sqrt{2N_\mathbf{u}+D-2}}\ri) \frac{1}{s^2}\,\le(s-N_\mathbf{u}\ri)\, \bigg]~\mathbf{\Phi}_s(x,\mathbf{u})=0\,.
\label{MdD omega 1}
\eea
At the end, by taking the limit ($m\rightarrow 0$\,, $s \rightarrow \infty$\,, $ms=\m=\mbox{constant}$) from the later, conveniently, we will arrive at the modified de Donder gauge condition for the bosonic continuous spin gauge field \eqref{MdD gauge22}\, and correspondingly to the modified CSP action \eqref{action}\,.

\subsection{Massive fermionic higher-spin action}

The gauge invariant action of massive arbitrary spin fields in D-dimensional flat and (A)dS$_D$ spacetimes were first found in \cite{Metsaev:2006zy}\,. The kinetic operator consists of two parts, a part depending on derivative and one depending on mass and spin\,. In flat spacetime, the derivative depending part corresponds to the terms appearing in the first two lines of the fermionic operator \eqref{eom F 3}\,. This part is independent of mass and spin\,, and thus will remain unchanged in the limit\,. Hence, we will here consider the part depending on the mass $m$ and half-integer spin $s=n+\frac{1}{2}$\,, with $n$ as a positive integer number\,. This part is given by \cite{Metsaev:2006zy}
\bea
\mkern-40mu {\mathcal{M}}_{\,(m,s)}&=&
 ~m\,\le[\,1\,-\,(\g\cdot {u})(\g\cdot\p_{u})\,-\,\frac{1}{4}\,{u}^2\,(\p_{u}\cdot\p_{u}) \, \ri] ~\frac{2n+D-2}{\,2n+D-2-2N_v\,}\nonumber \\
&+& ~ m\,\le[\,\g \cdot {u} \,-\,\frac{1}{2}\,{u}^2\,(\g\cdot \p_{u})\,\ri]\,\le(\frac{2n+D-3-N_v}{2n+D-4-2N_v} \ri)^{\frac{1}{2}} \,\p_v\nonumber \\
&-& ~ m\,v\,\le(\frac{2n+D-3-N_v}{2n+D-4-2N_v} \ri)^{\frac{1}{2}}\,\le[\,\g \cdot \p_{u} \,-\,\frac{1}{2}\,(\g\cdot {u})(\p_{u}\cdot\p_{u})\,\ri]
\,.  \label{eom F 2 m}
\eea
%
These operators act on the massive fermionic gauge fields $\psi_n(x,u,v)$\,, introduced by a same generating function as \eqref{varphi}\,. By following the procedure in the bosonic case, the singular factor can be extracted by introducing $\psi_n(x,u,v)=\a^n\,\mathbf{\Psi}_n(x,\w)$\,. Then, the well defined operator would be
\bea
\mkern-40mu {\mathbf{M}}_{\,(m,s)}&=&
 ~\m\,\le[\,1\,-\,(\g\cdot \mathbf{u})(\g\cdot\p_\mathbf{u})\,-\,\frac{1}{4}\,\mathbf{u}^2\,(\p_\mathbf{u}\cdot\p_\mathbf{u}) \, \ri] ~\frac{1}{\,n\,}~\frac{2n+D-2}{\,2N_\mathbf{u}+D-2\,}\nonumber \\
&+& ~ \m\,\le[\,\g \cdot \mathbf{u} \,-\,\frac{1}{2}\,\mathbf{u}^2\,(\g\cdot \p_\mathbf{u})\,\ri] ~\frac{\,N_\mathbf{u}+D-2+n\,}{\sqrt{2N_\mathbf{u}+D-2\,}} \,\frac{1}{\,n^2\,}\,(n-N_\mathbf{u}) \nonumber \\
&-& ~ \m~\frac{1}{\sqrt{2N_\mathbf{u}+D-2}}~\le[\,\g \cdot \p_\mathbf{u} \,-\,\frac{1}{2}\,(\g\cdot \mathbf{u})(\p_\mathbf{u}\cdot\p_\mathbf{u})\,\ri]
\,,  \label{eom F 3 m}
\eea
where, here, the applied change of variable is
\be
\w^{\a}\,=\,\mathbf{u}^{\,\a}~{\sqrt{{N_\mathbf{u}}+D-2+n}}\,.
\ee
If we now take into account the infinite spin limit of \eqref{eom F 3 m}\,, accompany with the derivative depending part, we will reach to the alternative operator for \eqref{eom F 3}\,(by substituting $\m \rightarrow -\,\m$ in \eqref{eom F 3})\,, and likewise to the modified fermionic CSP action \eqref{action F}\,.

A similar procedure, as above, can be carried out on the massive bosonic and fermionic actions in (A)dS spacetime (refs. \cite{Metsaev:2014iwa}\,, \cite{Metsaev:2006zy}) and compare the results with the proposed modified CSP actions in \cite{Metsaev: B CSP} and \cite{Metsaev: F CSP}\,.


\section{Conclusions and future directions}\label{s8}

In this paper, we presented the modified Wigner equations for both kinds of CSPs, bosons and fermions, which had not been discussed already in the literature\,. The idea to discover such equations was finding equations which can be obtained from the gauge-fixed modified CSP equations of motion\,. By focusing on the second bosonic (fermionic) Wigner equation, we redefined the gauge field such that the (gamma) trace-like condition converted to the (gamma) trace one\,. Indeed, for instance, the equations \eqref{cspin2} and \eqref{cspin3}\,, expressing that a CSP field is trace-full and divergence-free, reduced to the equations \eqref{cspin22} and \eqref{cspin33}\,, stating that a CSP field is trace-free and divergence-full\,! However, this is not surprising because, by a field redefinition, there might be found a set of modified higher-spin equations for fields which are trace-full and divergence-full\,. Moreover, by another kind of field redefinition, there might exist a system of exotic Wigner equations for an exotic field which is trace-full and divergence-full, i.e. when $0< \sigma < 1$\,. Therefore, depending on the problem of interest, for studying some problems the choice of Wigner equations can be more convenient while for studying other problems the use of modified Wigner equations will be preferable.

In the Fronsdal-like equation, a similar field redefinition was applied to transform the double trace-like constraint to the double trace one\,. Then, using the redefined field and applying a change of auxiliary variable, we arrived at the modified CSP action in D dimensions which was precisely the flat spacetime limit of the Metsaev action \cite{Metsaev: B CSP}\,. Exploring this outcome was particularly interesting and main goal of this paper\,. The gauge symmetry of the action was also found\,. By fixing a proper gauge in the modified equation of motion, we obtained the proposed modified Wigner equations, presented in the section \ref{MWE}\,. Further, we tried to find some relations to the Schuster-Toro action \cite{ST PRD}\,. For this purpose, we could establish a close relationship between the introduced operators ($\mathbf{P}_\Phi$\,, $\mathbf{P}_\varepsilon$) of this paper, used for rearranging the traces, and the obtained generating functions ($G(\w)$\,, $G'(\w)$) of \cite{ST PRD}, employed for acquiring the constrained formalism of the CSP action using a gauge fixing\,. In addition, by applying another change of auxiliary variable, we found a tensor form of the modified CSP action in D dimensions which is indeed the tensor action of \cite{ST PRD} in 4 dimensions\,.

A similar procedure was carried out on the Fang-Fronsdal-like equation, where the triple gamma trace-like constraint converted to the triple gamma trace one, and led to the modified fermionic CSP action in flat spacetime\,. The gauge symmetry and the gauge-fixed equation of motion (leading to the modified Wigner equations) were also conveniently found. We did not pursue finding a tensor form of the modified CSP action for fermions in this paper, however, it would be interesting to be explored by applying the similar manner of the bosonic case \cite{ST PRD} for the unconstrained fermionic action \cite{BNS} or using the way in the subsection \ref{S-T points} for the modified fermionic CSP action \eqref{action F}\,.

Finally, following the procedure of \cite{BM}, we explained how to obtain the modified CSP actions \cite{Metsaev: B CSP, Metsaev: F CSP} in flat spacetime from the massive higher spin actions \cite{Metsaev:2008fs, Metsaev:2006zy}\, by taking the infinite spin limit. This issue will make significant the study of the massive higher spin fields since they can be led, in a limit, to the continuous spin field\,.

The current exchanges for the constrained and unconstrained higher spin gauge fields were obtained in \cite{FMS} and it was shown that these results are equal to each other\,. Since the higher spin theory is a limit of the continuous spin theory, it would be nice to examine this fact for the constrained and unconstrained CSP fields\,. Considering the unconstrained gauge fields appeared in the Schuster-Toro action, the current exchanges mediated by a continuous spin particle have been investigated in \cite{BMN}\,. For future direction, it may be interesting to probe the current exchanges by taking into account the constrained gauge fields, appearing in the Metsaev action\,. It would be satisfactory to find a similar result as what was found in \cite{BMN}\,.

\paragraph{Note added:} While this work was in preparation for publication, a number of papers \cite{Bekaert:2017khg}-\cite{Schroer:2017nly} appeared in the context of the continuous (infinite) spin gauge theory\,.

\acknowledgments

We are extremely grateful to Xavier Bekaert for many helpful discussions and exchanges, without which this work could not be completed\,. We also thank Mohammad Khorrami for discussions and Ruslan Metsaev for comments on an earlier draft of the paper which led to a number of improvements. The author is also grateful to Jihad Mourad for collaboration on continuous spin particle, to Mohammad Mehdi Sheikh-jabbari, Mohammad Reza Setare and Hamid Reza Afshar for their support and encouragement, and to the LMPT and IPM members for their kind hospitality\,.

\appendix


\section{Conventions} \label{conv.}
We use the mostly plus signature for the metric and work in the D dimensional Minkowski spacetime\,. The convention
\be
\p_{\,\w_{\,\n}}=\frac{\p}{\p\,{\w^{\,\n}}}\,, \quad\quad\quad   \n=0,1,\ldots,D-1\,,
\ee
and the following commutation relations
\be
\le[\,\p_{\omega }^{\,\a} \,,\, \omega^{\,\b}\,\ri] =  \eta^{\,\a\b}\,,  \label{commutation 1}
\quad \quad \quad \quad
\le[\,{\dw\cdot\dw} \,,\, \omega^{\,2}\,\ri] =  4\, (N_\w + \tfrac{\,D\,}{2}) \,, \quad  N_\w= \omega \cdot \p_\omega\,,
\ee
are used\,. The hermitian conjugates introduce as
\be
(\p_x^{\,\a})^\dag:=\,-\,\p_x^{\,\a} \,,\qquad (\dw^{\,\a})^\dag := \w^{\,\a} \,,\qquad (\w^{\,\a})^\dag:=\dw^{\,\a} \,.
\label{hermitian conjugates}
\ee
%
For the D-dimensional Dirac gamma-matrices we use
the conventions
\be
\big\{\,\g^\a \,,\,\g^\b \,\big\}=2\,\e^{\,\a\b}
\,,
\quad\quad
(\,\g^\a\,)^{\,\dag} = \g^0\,\g^\a\,\g^0 \,,
\quad\quad
(\,\g^0\,)^{\,\dag} = -\,\g^0\,,
\quad\quad
(\,\g^0\,)^{\,2} = -\,1\,,
\label{anti comm gamma}
\ee
\be
\big\{\,\gamma \cdot \p_\omega \,,\, \gamma \cdot \omega \,\big\} =2\,(N_\w + \tfrac{\,D\,}{2})
\,.
\ee

\section{Transformation operators}\label{Trans. oper.}

In this appendix, we present operators of $\mathbf{P}_\varepsilon$ and $\mathbf{P}_\Phi$ which transform the Fronsdal-like constraints to the Fronsdal ones, and operators of $\mathbf{P}_\xi$ and $\mathbf{P}_\Psi$ which convert the Fang-Fronsdal-like constraints to the Fang-Fronsdal ones\,. We present these operators separately in the following.


\hfill

\noindent\textbf{$\mathbf{P}_\varepsilon$ operator:}

The Fronsdal-like equation \eqref{eom tilde} is invariant under the gauge transformation \eqref{Gauge T} so that the gauge transformation parameter obeys the ``trace-like constraint''
\be
( \p_\omega \cdot \p_\omega +\,\sigma ) \, \widetilde{\varepsilon}\,(x,\omega) =0 \label{shifted eps} \,,
\ee
where the parameter $\s$ was introduced in \eqref{sigma}\,.
Then, we introduce the operators \footnote{\,The author is grateful to Xavier Bekaert for helpful discussions leading to these operators and corresponding ones\,.
}
\be
\mathbf{P}_\varepsilon = ~ \sum_{n=0}^{\infty}~\sigma^n~ \omega^{\,2n}~ \frac{(-1)^n}{~2^{\,2n} ~ n!~ (N+\tfrac{D}{2})_n~} \,, \label{P_eps}
\ee
and
\be
~~~~\,\mathbf{P}'_\varepsilon =~ \sum_{n=0}^{\infty}~\sigma^n~\omega^{\,2n}~ \frac{(-1)^n}{~2^{\,2n} ~ n!~ (N+\tfrac{D}{2}+2)_n~} \,,
\label{P'_eps}
\ee
where D is the dimension of spacetime, $N={\w\cdot\dw}$\, and $(a)_n$ is the rising Pochhammer symbol \eqref{Pochhammer}\,. Note that the operator $\mathbf{P}'_\varepsilon$ is nothing except the operator $\mathbf{P}_\varepsilon$ in which its N is shifted to N+2\,.
Now using \eqref{P_eps}\, and \eqref{P'_eps}\,, it is convenient to illustrate
\be
\mathbf{P}'_\varepsilon ~( \p_\omega \cdot \p_\omega ) = ( \p_\omega \cdot \p_\omega +\,\sigma )~\mathbf{P}_\varepsilon\,. \label{P'=P}
\ee
To obtain this relation, we use the relations \eqref{21}\, and \eqref{poch 2}\,. Note that $\mathbf{P}_\varepsilon=1=\mathbf{P}'_\varepsilon$\, when $\sigma=0$\,. Using the operator $\mathbf{P}_\varepsilon$\,, we can write $\widetilde{\varepsilon}$ in terms of a new gauge transformation parameter $\varepsilon$
\be
\widetilde{\varepsilon}\,(x,\omega) = \mathbf{P}_\varepsilon ~ \varepsilon(x,\omega)\,.
\label{redif eps}
\ee
Plugging \eqref{redif eps} into \eqref{shifted eps} and then using \eqref{P'=P}\,, we can display
\be
( \p_\omega \cdot \p_\omega )\, \varepsilon(x,\omega) =0\,,
\ee
which is the ``trace constraint'' of the new gauge transformation parameter\,.


\hfill

\noindent\textbf{$\mathbf{P}_\Phi$ operator:}

In the Fronsdal-like equation \eqref{eom tilde}\,, the ``double trace-like constraint'' on the gauge field is given by
\be
\le( \,\p_\omega \cdot \p_\omega +\,\sigma \,\ri)^2 ~ \widetilde{\Phi}\,(x,\omega) =0\,.
\ee
Introducing operators
\be
\mathbf{P}_\Phi = ~ \sum_{n=0}^{\infty}~ \sigma^n~\omega^{\,2n}~ \frac{(-1)^n}{~2^{\,2n} ~ n!~ (N+\tfrac{D}{2}-1)_n~} \,, \label{P_phi}
\ee
and
\be
\mathbf{P}'_\Phi =~\sum_{n=0}^{\infty}~\sigma^n~\omega^{\,2n}~ \frac{(-1)^n}{~2^{\,2n} ~ n!~ (N+\tfrac{D}{2}+3)_n~} \,,
\ee
we can show
\be
\mathbf{P}'_\Phi ~( \p_\omega \cdot \p_\omega )^2 = ( \p_\omega \cdot \p_\omega +\,\sigma )^2~\mathbf{P}_\Phi\,.
\ee
To get the later, we use the commutation relation \eqref{41} and the properties of the Pochhammer symbol\,. Then, using the $ \mathbf{P}_\Phi$ operator, we can introduce
\be
\widetilde{\Phi}\,(x,\omega) = \mathbf{P}_\Phi ~ \Phi(x,\omega)\,,
\ee
where $\Phi$ is a new gauge field satisfying the ``double trace constraint''
\be
(\, \p_\omega \cdot \p_\omega\, )^{\,2}~ \Phi(x,\omega) =0\,.
\ee
Note that, the operators $\mathbf{P}_\varepsilon$ and $\mathbf{P}_\Phi$ are related to each other via
\be
\w^{\,\n}~\mathbf{P}_\varepsilon = \mathbf{P}_\Phi~\w^{\,\n}\,.
\label{Peps - Pphi}
\ee
Moreover, applying \eqref{poch 1} in $\mathbf{P}_\Phi$ we can show another useful relation
\be
\mathbf{P}_\varepsilon \,=\,\mathbf{P}_\Phi~\le(\,1\,+\,\s\, \w^2 ~ \frac{1}{\,(2N+D)(2N+D-2)\,}  \ri)\,.
\label{Peps - Pphi 1}
\ee


\hfill

\noindent\textbf{$\mathbf{P}_\xi$ operator:}

The Fang-Fronsdal-like equation \eqref{eom tilde F} is invariant under the gauge transformation \eqref{Gauge T F} with the ``gamma-trace-like constraint'' on the gauge transformation parameter
\be
\left(\, \g \cdot \p_\omega \,+\,i\, \sigma\, \right) \, \widetilde{\xi}\, (x,\omega) = 0\,.   \label{xi constraint 1}
\ee
Introducing the operators
\be
\mkern-30mu\mathbf{P}_\xi =\sum_{k=0}^{\infty}\le[(\sigma\,\gamma \cdot \omega)^{2k} + 2ik\, (\sigma\,\gamma \cdot \omega)^{2k-1}\ri]
\frac{(-1)^k}{~2^{\,2k}~k!~(N+\tfrac{D}{2})_k~}  \,,
\label{P xi}
\ee
and
\be
\mathbf{P}'_\xi =\sum_{k=0}^{\infty}\le[(\sigma\,\gamma \cdot \omega)^{2k} - 2ik\, (\sigma\,\gamma \cdot \omega)^{2k-1}\ri]
\frac{(-1)^k}{~2^{\,2k}~k!~(N+\tfrac{D}{2}+1)_k~}\,,
\label{P' xi}
\ee
we can prove
\be
\mathbf{P}'_\xi \,\le(\gamma \cdot \p_\omega\ri) = \le(\, \gamma \cdot \p_\omega \,+ i\,\sigma\,\ri) \mathbf{P}_\xi \,.
\label{P' xi to xi}
\ee
To acquire the later, we use the commutation and anti-commutation relations \eqref{11}\,, \eqref{12}\,. Introducing a new gauge transformation parameter $\xi$
\be
\widetilde{\xi}(x,\omega) =\mathbf{P}_\xi ~\xi(x,\omega)\,,
\ee
we find the ``gamma-trace constraint'' on $\xi$ as
\be
(\,\g\cdot\dw\,)~\xi(x,\omega)=0\,.
\ee


\hfill

\noindent\textbf{$\mathbf{P}_\Psi$ operator:}

In the Fang-Fronsdal-like equation \eqref{eom tilde F}\,, the ``triple gamma-trace-like constraint'' on the gauge field
\be
\le(\g \cdot \dw + i\,\s \ri)\le(\dw \cdot \dw +\,\s \ri)\widetilde{\Psi}(x,\omega) =0\,, \label{Field cons. F 1}
\ee
can be reduced to the ``triple gamma-trace constraint'' on the new gauge field
\be
\le(\g \cdot \dw \ri)^3 {\Psi}(x,\omega) =0\,, \label{Field cons. F22}
\ee
such that
\be
\widetilde{\Psi}(x,\omega) =\mathbf{P}_\Psi ~\Psi(x,\omega)\,, \label{psi' psi}
\ee
and
\be
\mathbf{P}_\Psi =\sum_{k=0}^{\infty}\le[(\sigma\,\gamma \cdot \omega)^{2k} + 2ik\, (\sigma\,\gamma \cdot \omega)^{2k-1}\ri]
\frac{(-1)^k}{~2^{\,2k}~k!~(N+\tfrac{D}{2}-1)_k~} \,.
\label{P say}
\ee
To obtain \eqref{Field cons. F22}\,, we can illustrate
\be
\mathbf{P}'_\Psi \,\le(\gamma \cdot \p_\omega\ri)^3 = \le(\g \cdot \dw + i\,\s \ri)\le(\dw \cdot \dw +\,\s\, \ri) \mathbf{P}_\Psi \,, \label{p ksi' p ksi}
\ee
where
\be
~~\mkern-15mu\mathbf{P}'_\Psi =\sum_{k=0}^{\infty}\le[(\sigma\,\gamma \cdot \omega)^{2k} - 2ik\, (\sigma\,\gamma \cdot \omega)^{2k-1}\ri]
\frac{(-1)^k}{~2^{\,2k}~k!~(N+\tfrac{D}{2}+2)_k~}\,.
\ee
To show \eqref{p ksi' p ksi}\,, the commutation and anti-commutation relations \eqref{31}\,, \eqref{32} are used\,. Similar to the bosonic case, it can be shown that the fermionic operators $\mathbf{P}_\xi$ and $\mathbf{P}_\Psi$ are related to each other via
\be
\w^{\,\n}~\mathbf{P}_\xi = \mathbf{P}_\Psi~\w^{\,\n}\,,
\label{Pkesi--Ppsi}
\ee
\be
\mathbf{P}_\xi = \mathbf{P}_\Psi \le(\,1\,+\,i\,\sigma\,(\gamma \cdot \omega) ~\frac{2}{(2N+D)(2N+D-2)}\,+\,\sigma\,\w^2~ \frac{1}{(2N+D)^2} \ri)\,.
\label{Pkesi--Ppsi 1}
\ee

\section{Modified Wigner equations} \label{MWEss}
In this appendix, using the results in the previous section, we will explain how to get the modified Wigner equations in \eqref{cspin11}-\eqref{cspin44} and \eqref{fcspin11}-\eqref{fcspin44}\,, from the bosonic and fermionic Wigner equations\,.

\hfill

\noindent\textbf{Bosonic equations:}

The first modified Wigner equation \eqref{cspin11} can be obtained by plugging \eqref{P_eps 1} into \eqref{cspin1}\,, and then\,, by applying \eqref{Peps - Pphi} and \eqref{Peps - Pphi 1}\,. The second equation \eqref{cspin22} will obtain from \eqref{P'=P}\,. To obtain the third equation \eqref{cspin33}\,, we first show
\bea
\mkern-100mu && \dw^{\,\a}~ \mathbf{P}_\epsilon = \mathbf{P}_\Phi ~\Big[\,\dw^{\,\a} -\,\s\,\frac{1}{2N+D-2}\,\w^\a + \,\s\,
\w^2\, \frac{2}{(2N+D-2)(2N+D+2)}\,\dw^{\,\a} \nonumber \\
\mkern-100mu&& \quad\quad\quad\quad\quad\quad\quad\,~\, - \,\s\, \w^2 \,\frac{1}{(2N+D)(2N+D-2)^2}\,\w^\a \nonumber\\
\mkern-100mu&& \quad\quad\quad\quad\quad\quad\quad\,~\, + \,\s\,\w^4\, \frac{1}{(2N+D)(2N+D+4)(2N+D+2)^2}\,\dw^{\,\a}\, \Big]\,.
\label{dw P epsi}
\eea
Then by applying \eqref{P_eps 1} in \eqref{cspin3}\,, and using the later\,, we will reach to
\bea
\mkern-50mu\mathbf{P}_\Phi ~\Big[\Big(\,1+\,\s\,
\w^2\, \tfrac{2}{(2N+D-2)(2N+D+2)}\,
+ \,\s\,\w^4\, \tfrac{1}{(2N+D)(2N+D+4)(2N+D+2)^2}\,\Big)\big(p\cdot\dw\big)  \\
\qquad-\,\Big(\tfrac{1}{2N+D-2}\,+\,\s\, \w^2 \,\tfrac{1}{(2N+D)(2N+D-2)^2} \Big)\big(p\cdot\w\big)\Big] \nonumber \boldsymbol{\varphi}(p,\w)&=&0\,.
\eea
This relation can be simplified (by applying the first modified Wigner equation \eqref{cspin11}) to
\be
\mathbf{P}_\Phi \,\Big[1+
\w^2\, \tfrac{2\,\s}{(2N+D-2)(2N+D+2)}
+ \w^4\, \tfrac{\s}{(2N+D)(2N+D+4)(2N+D+2)^2}\Big]\Big[\,{p\cdot\dw}+\tfrac{\s\,\m}{2N+D-2}\, \Big]\boldsymbol{\varphi}=0\,,
\ee
such that the third modified Wigner equation \eqref{cspin33} can be conveniently read of it\,.

\hfill

\noindent\textbf{Fermionic equations:}

Putting \eqref{P Kesi} in \eqref{fcspin1}\,, and then using \eqref{Pkesi--Ppsi} and \eqref{Pkesi--Ppsi 1}\,, we will get the first modified Wigner equation \eqref{fcspin11}\,. To get the second equation \eqref{fcspin22}\,, using \eqref{P Kesi}\,,  we can write \eqref{fcspin2} as
\be
\le(\, \gamma \cdot \p_\omega \,- i\,\sigma\,\ri)\le(\, \gamma \cdot \p_\omega \,+ i\,\sigma\,\ri) \mathbf{P}_\xi~ \boldsymbol{\psi}(p,\w)=\le(\, \gamma \cdot \p_\omega \,- i\,\sigma\,\ri)\mathbf{P'}_\xi\,\le(\, \gamma \cdot \p_\omega\ri) \, \boldsymbol{\psi}(p,\w)=0\,,
\label{P'xi-P'epsi}
\ee
which the relation \eqref{P' xi to xi} is used\,.
Then, by illustrating
\be
\le(\, \gamma \cdot \p_\omega \,- i\,\sigma\,\ri)\mathbf{P'}_\xi = \mathbf{P'}_\varepsilon \le({\g\cdot\dw} - \,i\,\s\,(\g\cdot\w)(\g\cdot\dw)\,\tfrac{1}{2N+D+2}\,-\,i\,\s\,\tfrac{2}{2N+D+2} \ri)\,,
\ee
and applying it in \eqref{P'xi-P'epsi}\,, we will get the second equation \eqref{fcspin22}\,. Looking at the form of the operators \eqref{P_eps} and \eqref{P xi}\,, we can show
\be
\mathbf{P}_\xi = \mathbf{P}_\varepsilon \le( 1 -\,i\,\sigma \,(\gamma \cdot \omega) ~\frac{1}{2N+D} \ri)\,,
\ee
and then using \eqref{dw P epsi}\,, and the way was applied to get the correspondence bosonic equation\,, we will obtain the third modified equation \eqref{fcspin33}\,. Applying \eqref{P Kesi} in \eqref{fcspin4} we will find
\be
\le(\g\cdot p\ri)\,\mathbf{P}_\xi~ \boldsymbol{\psi}(p,\w) = \mathbf{P}_\varepsilon \le(\g\cdot p\ri) \le( 1 -\,i\,\sigma \,(\gamma \cdot \omega) ~\frac{1}{2N+D} \ri)\,\boldsymbol{\psi}(p,\w) =0 \,,
\ee
or (using the anti-commutation relation in \eqref{anti comm gamma})
\be
\le[\,{\g\cdot p}\, + \,i\,\s\,(\g\cdot\w)(\g\cdot p)\,\frac{1}{2N+D}\, -\,i\,\s\,\frac{2}{2N+D-2}\,(p\cdot\w)\, \ri]\boldsymbol{\psi}(p,\w) =0\,.
\label{gamma.p}
\ee
Afterward, applying the first modified equation \eqref{fcspin11}\,, the equation \eqref{gamma.p} will be converted to
\be
\le[1+\s\,\w^2\,\tfrac{1}{(2N+D)(2N+D+2)} \ri]\le[\gamma \cdot p \,+\,i\,\s\,\m\,\tfrac{2}{2N+D-2}\,+\,\s\,\m\,(\g\cdot\w)\,\tfrac{2}{(2N+D)^2}\, \ri]\boldsymbol{\psi}(p,\w) =0\,,
\ee
where the forth modified Wigner equation \eqref{fcspin44} will be appeared\,.

\section{Useful relations}\label{useful}
This appendix contains useful relations which are used and computed for the text\,.

The \textit{Bessel function of the first kind} $J_\a(z)$ is given as the power series
 \be
 J_\a(z) \, = \, \sum_{m=0}^{\infty} \, \frac{(-1)^m}{ \, m! \, \Gamma(m + \a +1)} ~ \le( \frac{z}{2} \ri)^{2m+\a} \,. \label{Bessel}
 \ee

The \textit{rising Pochhammer symbol} $(a)_n$ is defined as
\be
 (a)_n = 
a \, (a+1)(a+2) \cdots (a+n-1) = \frac{\Gamma(a+n)}{\Gamma(a)} \,, \quad\quad n\in\mathbb N ~~\mbox{and}~~ a\in\mathbb R\,.  \label{Pochhammer}
 \ee
Then, the following useful properties can be obtained
\be
\frac{~a+n-1~}{~(a)_n~}~=~\frac{a-1}{~(a-1)_n~}\,,
\label{poch 1}
\ee

\be
(a)_n ~=~ (a+n-1)~ (a)_{n-1}\,,
\label{poch 2}
\ee

\be
(a)_n ~=~a~ (a+1)_{n-1}\,.
\label{poch 3}
\ee

The useful commutation (\,$ [ a , b ] = ab - ba $\,) and anti-commutation (\,$\{ a , b \} = ab + ba $\,) relations, for any integer $k \geqslant 1$\,, have been computed as follow:
\bea
\mkern-50mu
\le[\,\gamma \cdot \p_\omega ~, ~(\gamma \cdot \omega)^{\,2k} \,\ri] &=&2k\, (\gamma \cdot \omega)^{\,2k-1} \,,  \label{11} \\
\mkern-50mu
\le\{\gamma \cdot \p_\omega \,,\, (\gamma \cdot \omega)^{\,2k-1} \ri\} &=&2(N+\tfrac{D}{2}-k+1)~ (\gamma \cdot \omega)^{\,2k-2} \,,
\label{12}
\eea

\bea
\mkern-50mu\le[\,(\dw \cdot \p_\omega) ~, ~ \omega^{\,2k} \,\ri] &=&4k(N+\tfrac{D}{2}-k+1)~  \omega^{\,2k-2} \,,
 \label{21}
\eea

\bea
\mkern-2mu\le[\,(\gamma \cdot \p_\omega)^3 ~, ~(\gamma \cdot \omega)^{\,2k} \,\ri] &=&2k(\gamma \cdot \p_\omega)^2 (\gamma \cdot \omega)^{2k-1}\,+ 4k(N+\tfrac{D}{2}-k+1) \times \label{31} \\
&\times &  \le[ (\gamma \cdot \p_\omega)(\gamma \cdot \omega)^{\,2k-2} -2(k-1)(\gamma \cdot \omega)^{\,2k-3} \ri]\,, \nonumber \\
%
\mkern-2mu
\le\{(\gamma \cdot \p_\omega)^3 , (\gamma \cdot \omega)^{\,2k-1} \ri\} &=&2(N+\tfrac{D}{2}-k+2)\Big[ (\gamma \cdot \p_\omega)^2 (\gamma \cdot \omega)^{\,2k-2}\,+\, 2(k-1)\times \label{32}\\
& \times&  (\gamma \cdot \p_\omega) (\gamma \cdot \omega)^{\,2k-3}\,  -\,4(k-1)(N+\tfrac{D}{2}-k+1)(\gamma \cdot \omega)^{\,2k-4}  \Big]\,,\nonumber
\eea

\bea
\mkern-50mu\le[\,(\dw \cdot \dw)^{\,2} ~, ~\w^{\,2k} \,\ri] &=&8k(N+\tfrac{D}{2}-k+2)~ \Big[(\dw\cdot\dw)~\w^{\,2k-2}\,\nonumber\\
&& \quad\quad\quad\quad\quad\quad\quad\quad\quad~-\,2(k-1)(N+\tfrac{D}{2}-k+1)\,\w^{\,2k-4} \Big] \,.
 \label{41}
\eea


The act of quantities $\dw^{\,\a}$\,, $\dw^{\,2}$\, and $\w^\a$\, on the bosonic operator $\mathbf{P}_\Phi$\,, introduced in \eqref{P_phi}\,, have computed as
\be
~~\dw^{\,\a}~\mathbf{P}_\Phi=\mathbf{P}_\Phi\bigg[\, \dw^{\,\a}\,+\,\w^2~\frac{\sigma}{(2N+D)(2N+D-2)}~\dw^{\,\a}\,-\,\w^\a~\frac{\sigma}{(2N+D-2)}\,\bigg]\,,  \label{1}
\ee
\bea
\mkern-40mu\dw^{\,2}~\mathbf{P}_\Phi&=&\mathbf{P}_\Phi\bigg[\,(\dw\cdot\dw)\,+\,\w^2~\frac{2\,\sigma}{(2N+D-2)(2N+D+2)}~(\dw\cdot\dw)
\,-\,\sigma\nonumber \\
&&\quad~~~-\,\frac{2\,\sigma}{(2N+D-2)} \,-\,\w^2~\frac{2\,\sigma}{(2N+D)(2N+D-2)^2}+~  \mathcal{O}(\w^4) \bigg]\,, \label{2}
\eea
\be
\mkern-155mu~~~\,\w^\a~\mathbf{P}_\Phi=\mathbf{P}_\Phi\bigg[\,\w^\a\,-\,\w^2\, \w^\a~\frac{\sigma}{(2N+D)(2N+D-2)}~+~  \mathcal{O}(\w^4) \bigg]\,, \label{3}
\ee
where the terms containing $\mathcal{O}(\w^4)$ in the last two relations, will be omitted at the level of the action, due to the double trace constraint on the gauge field ${\Phi}(x,\dw)\,(\w^2)^{\,2}\,=0$\,.
%

On the other hand, the act of ${\g\cdot\p_x}$\,, ${\w\cdot\p_x}$ and ${\gamma \cdot \dw}$ on the fermionic operator $\mathbf{P}_\Psi$\,, given by \eqref{P say}\,, have computed as
\bea
\mkern-30mu({\gamma \cdot \p_x})~ \mathbf{P}_\Psi&=&\mathbf{P}_\Psi\bigg[\, ({\gamma \cdot \p_x})\,-\,i\,({\w\cdot\p_x})~\frac{2\,\sigma}{(2N+D-2)}\,+\,i\,({\gamma\cdot\w})({\gamma \cdot \p_x})~\frac{2\,\sigma}{(2N+D-2)}\nonumber\\
&&\quad\quad\quad\quad\quad~ +\,({\gamma\cdot\w})({\w\cdot\p_x})~\frac{2\,\sigma}{(2N+D)(2N+D-2)} \nonumber\\
&&\quad\quad\quad\quad\quad~ -~\w^2\,({\gamma \cdot \p_x})~\frac{2\,\sigma}{(2N+D)(2N+D-2)}~+~  \mathcal{O}(\w^3) ~ \bigg]\,,  \label{1 1}
\eea
\be
\mkern-10mu\,({\w\cdot\p_x})~\mathbf{P}_\Psi=\mathbf{P}_\Psi\bigg[\,({\w\cdot\p_x})
\,-\,i\,({\gamma\cdot\w})({\w\cdot\p_x})~\frac{2\,\sigma}{(2N+D)(2N+D-2)}~+~  \mathcal{O}(\w^3)~ \bigg]\,, \label{2 2}
\ee
\bea
\mkern-56mu({\gamma \cdot \dw})~ \mathbf{P}_\Psi&=&\mathbf{P}_\Psi\bigg[\, ({\gamma \cdot \dw})\,+\,
({\gamma \cdot \w})~\frac{2\,\sigma}{(2N+D-2)^2}  \label{3 3} \\
&&\quad\quad\quad\quad\quad~-\,\w^2~\frac{\sigma}{(2N+D)^2} ~ ({\gamma \cdot \dw})\, -\,i\,\s ~\frac{2N+D}{(2N+D-2)} \nonumber \\
&&\quad\quad\quad\quad\quad~+\,i\,\s\,({\gamma \cdot \w}) \,\frac{2(2N+D-1)}{(2N+D)(2N+D-2)}\,({\gamma \cdot \dw})
 \, +  \mathcal{O}(\w^3) \, \bigg]\,,  \nonumber
\eea
where the terms containing $\mathcal{O}(\w^3)$ will be vanished, at the level of the action, because of the triple gamma-trace constraint on the spinor gauge field ${\overline{\Psi}}(x,\dw)\,(\g\cdot\w)^3=0$\,.

%




\end{document}